\documentclass{article}

\usepackage{PRIMEarxiv}
\usepackage[utf8]{inputenc}
\usepackage[T1]{fontenc}
\usepackage{hyperref}
\usepackage[numbers,sort&compress]{natbib}
\usepackage{url}
\usepackage{booktabs}
\usepackage{nicefrac}
\usepackage{microtype}
\usepackage{fancyhdr}
\usepackage{float}
\usepackage[font=small,labelfont=bf]{caption}
\usepackage{enumitem}
\usepackage{graphicx}
\usepackage{subcaption}
\usepackage{xcolor}

\newif\ifshowrev
\showrevtrue
\ifshowrev
  \newcommand{\revcolor}{\color{blue}}
\else
  \newcommand{\revcolor}{}
\fi

\newcommand{\Cav}{C_{\mathrm{AV}}}
\newcommand{\mSparse}{\textsc{Sparse}}
\newcommand{\mMedium}{{\textsc{Medium}}}
\newcommand{\mDense}{{\textsc{Dense}}}

\newcommand{\Ctax}{C_{\mathrm{tax}}}
\newcommand{\psub}{p_{\mathrm{sub}}}

\usepackage{amsmath, amsfonts, amssymb, amsthm, mathtools}
\usepackage{bm}
\usepackage{multirow}
\usepackage{siunitx}
\usepackage{algorithm}
\usepackage{algpseudocode}
\usepackage{tikz}
\usepackage{tikz-cd}
\usepackage{pgfplots}
\pgfplotsset{compat=1.18}

\usepackage[nameinlink,noabbrev,capitalize]{cleveref}

\usepackage[acronym]{glossaries-extra}
\setabbreviationstyle[acronym]{long-short}
\glssetcategoryattribute{acronym}{nohyperfirst}{true}
\newacronym{acr:amod}{AMoD}{Autonomous Mobility-on-Demand}
\newacronym{acr:mod}{MoD}{Mobility-on-Demand}
\newacronym{acr:pt}{PT}{Public Transit}
\newacronym{acr:mcp}{MCP}{Municipality}
\newacronym{acr:od}{OD}{Origin-Destination}
\newacronym{acr:mnl}{MNL}{Multinomial Logit}
\newacronym{acr:kkt}{KKT}{Karush-Kuhn-Tucker}
\newacronym{acr:micp}{MICP}{Mixed Integer Convex Program}
\newacronym{acr:sca}{SCA}{Successive Convex Approximation}
\newacronym{acr:ibr}{IBR}{Iterative Best Response}
\newacronym{acr:osm}{OSM}{OpenStreetMap}
\newacronym{acr:av}{AV}{Autonomous Vehicles}
\newacronym{acr:p2p}{P2P}{Peer-to-Peer}

\theoremstyle{plain}
\newtheorem{theorem}{Theorem}[section]
\newtheorem{lemma}{Lemma}[section]
\newtheorem{proposition}{Proposition}[section]
\newtheorem{corollary}{Corollary}[section]

\theoremstyle{definition}
\newtheorem{definition}{Definition}[section] 
\newtheorem{assumption}{Assumption}[section]
\crefname{assumption}{Assumption}{Assumptions}
\Crefname{assumption}{Assumption}{Assumptions}

\theoremstyle{remark}
\newtheorem{remark}{Remark}[section]

\title{To Own or Not to Own? Endogenous Ownership in Peer-to-Peer Sharing of Private Autonomous Vehicles}

\author{  \textbf{Xinling Li}$^{*}$ \quad
  \textbf{Gioele Zardini}$^{*}$ \\
  $^{*}$Laboratory for Information \& Decision Systems, Massachusetts Institute of Technology, Cambridge, MA, USA \\
  \texttt{xinli831@mit.edu}, \texttt{gzardini@mit.edu}
}

\date{}

\begin{document}
\maketitle

\begin{abstract}
Autonomous driving allows private owners to release their vehicles to an on-demand platform and earn income when they do not need them. This possibility changes vehicle acquisition incentives, as rental income can encourage purchases, while access to shared vehicles can reduce the need for personal ownership. Understanding these competing effects is essential to assessing the ownership implications of private \glspl{acr:av} sharing. While existing models of private \gls{acr:av} sharing often take ownership as given, in this paper, we develop a stationary equilibrium model in which vehicle ownership and listing decisions determine service availability and operational frictions, which in turn shape individuals’ incentives to own and share. Based on the model, we first characterize individual choices by comparing payoffs of being an owner and a non-owner. We then analyze the resulting potential population partitions, establish conditions for the existence of an active sharing market, and categorize them by owner types. The equilibrium is presented by a system of equations and solved through a tailored algorithm. An illustrative numerical study on calibrated U.S. travel and vehicle cost data shows that sharing can either increase or decrease ownership, depending on the balance between purchases encouraged by rental income and ownership displaced by shared vehicles. The balance is shaped by market parameters, including market density, \gls{acr:av} cost, and platform operating conditions. Platform pricing and municipal regulation can also influence ownership through their effects on acquisition incentives and the shared service that makes nonownership attractive. These findings highlight the range of possible ownership outcomes in private \gls{acr:av} sharing markets, and the model provides a framework for evaluating how market conditions, platform decisions, and regulation shape these outcomes.
\end{abstract}


\section{Introduction}
\glspl{acr:av} are moving from pilots to commercial service, and robotaxi fleets already operate in several cities, where ride-hailing platforms have started to integrate them into their networks~\citep{waymo2024uber}. While current discussion on ride-hailing with \gls{acr:av} often focuses on services operated with a centrally owned \gls{acr:av} fleet (e.g., Waymo and Zoox), a less discussed consequence of this transition concerns private cars.
American spend on average 58.6 minutes per day traveling by vehicle~\citep{mcguckin2018summary}, so a privately owned vehicle is parked for the vast majority of the day.
Today, this idle time can only be shared through \gls{acr:p2p} carsharing platforms such as Turo in the United States and GoMore in Europe, which require a physical handover and typically serve rentals of hours to days~\citep{dill2019effects,schwieterman2020estimating}.
Once a vehicle can drive itself, its owner can instead release it to an on-demand platform between their own trips and recall it when needed~\citep{stocker2017shared}.
Manufacturers are preparing this step. Tesla has announced that it aims to let customers add their own vehicles to its robotaxi network~\citep{fortune2025tesla,electrek2026tesla}, and it has begun inviting cuetomers to order its robotaxi~\citep{okane2026cybercab}.
When sharing income is comparable to the cost of a vehicle, this possibility affects how the existing private fleet is used and also who buys a vehicle in the first place.
The shared fleet of a city could then be owned by residents who share the spare time of the vehicle they use, or by residents who buy a vehicle only to share it, and which of the two groups owns the robotaxi shapes how many vehicles the city ends up with.

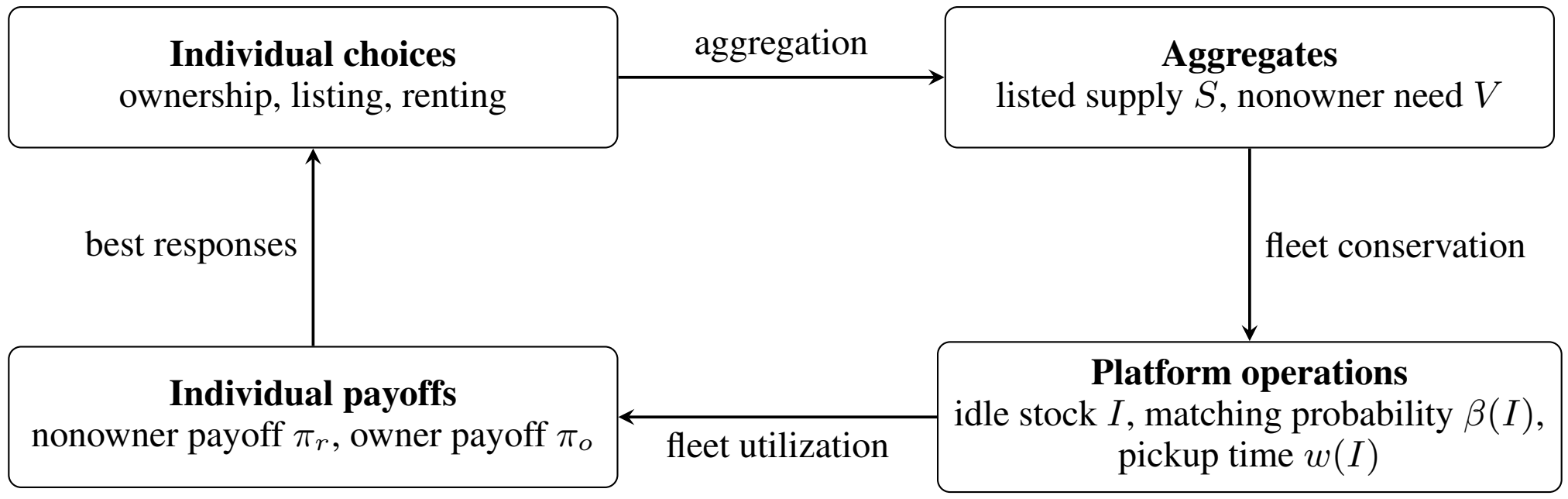
\begin{figure}[tb]
\centering
\begin{tikzpicture}[font=\small,
  box/.style={draw, rounded corners=3pt, align=center, minimum width=5.2cm, minimum height=1.2cm, inner sep=4pt},
  arr/.style={->, >=stealth, line width=.7pt}]
\node[box] (choice) at (0,0) {\textbf{Individual choices}\\ ownership, listing, renting};
\node[box] (agg) at (8,0) {\textbf{Aggregates}\\ listed supply $S$, nonowner need $V$};
\node[box] (ops) at (8,-2.9) {\textbf{Platform operations}\\ idle stock $I$, matching probability $\beta(I)$, \\ pickup time $w(I)$};
\node[box] (svc) at (0,-2.9) {\textbf{Individual payoffs}\\ nonowner payoff $\pi_r$, owner payoff $\pi_o$};
\draw[arr] (choice) -- node[above]{aggregation} (agg);
\draw[arr] (agg) -- node[right, align=left]{fleet conservation} (ops);
\draw[arr] (ops) -- node[below]{fleet utilization} (svc);
\draw[arr] (svc) -- node[left, align=right]{best responses} (choice);
\end{tikzpicture}
\caption{The equilibrium loop of a \gls{acr:p2p} market for private \glspl{acr:av}. Ownership, listing, and renting choices aggregate into listed supply and nonowner demand, which are operated by the platform, and operations determine the individual payoffs.}
\label{fig:loop}
\end{figure}

Understanding this market is difficult for three reasons.
First, ownership is endogenous. 
The option to share changes the economics of owning a vehicle, so the set of potential suppliers is itself an equilibrium outcome.
Second, a shared \gls{acr:av} differs from a durable good handed over for a fixed rental period.
Once listed, it is committed to trips whose timing the owner does not control, so the owner's access to their own vehicle depends on fleet utilization and on trip durations.
Symmetrically, the value of the service to a renter depends on whether an idle vehicle is available nearby and on how long the pickup takes.
Third, these effects close a loop.
Ownership and listing determine the idle stock, the idle stock determines service conditions, and the service conditions determine sharing income, access reliability, and the attractiveness of renting, which in turn affects ownership and listing decisions. This loop is illustrated in \Cref{fig:loop}.
Taken together, these obstacles call for a model in which ownership, listing, and platform operations are determined jointly in equilibrium.

Existing work addresses parts of this loop, and four families of approaches are relevant. First, the literature on the control of \gls{acr:amod} systems optimizes matching, rebalancing, pooling, and pricing for a fleet whose size is given or chosen by the operator~\citep{pavone2012robotic,zhang2016control,zardini2022analysis}. 
Second, work on mixed autonomy and flexible supply treats \glspl{acr:av} as committed capacity and human drivers as flexible capacity that responds to incentives~\citep{castro2024getting,freund2026supply,lian2025capturing}. 
Third, a growing literature studies privately owned \glspl{acr:av} as a decentralized source of capacity and models whether and how owners share~\citep{wang2021aggregate,siddiq2022ride,zhang2025park,wang2026sharing}. 
Finally, the economics literature on \gls{acr:p2p} product sharing shows that access to a rental market can raise or lower the ownership of durable goods~\citep{fraiberger2017peer,jiang2018collaborative,benjaafar2019peer,filippas2020owning}.

These works rely on assumptions that fail to capture the endogenous ownership effect of sharing with private \glspl{acr:av}.
Models of private \gls{acr:av} sharing typically assume that the population of \gls{acr:av} owners is fixed, so that the only endogenous decision is whether an existing vehicle is shared.
When sharing income is comparable to the cost of ownership, however, the option to share can induce new purchases or make existing ones unnecessary, and the fixed-ownership assumption can misrepresent both the size and the composition of the shared fleet.
Conversely, models of \gls{acr:p2p} product sharing endogenize ownership but treat a rental as the transfer of a good for a predetermined period, with matching frictions that do not depend on how the shared capacity is operated.
In on-demand mobility, by contrast, a renter's value depends on service conditions such as idle-vehicle availability and pickup times, and an owner's cost depends on whether the vehicle can be recalled in time.
This gap motivates a model that combines the endogenous ownership margin of \gls{acr:p2p} sharing with the operational frictions of \gls{acr:amod}.

Several questions then arise naturally. Under which conditions can an active sharing market emerge at all? When it does, does access to shared vehicles increase or reduce ownership? Answering this question requires understanding who owns, who shares, and who relies on the service instead of owning, and how market density, \gls{acr:av} acquisition costs, and platform operations shape these choices. The same tension raises questions for platforms and policymakers: are the platform's revenue incentives aligned with reducing vehicle ownership, and can a municipality discourage vehicle purchases without undermining the shared service that makes nonownership attractive? 
The purpose of this paper is to provide rigorous answers to these questions.

\paragraph*{Statement of contributions} This paper makes four contributions. First, we develop a stationary model of a \gls{acr:p2p} sharing market for private \glspl{acr:av} that combines endogenous ownership with the operational frictions of on-demand mobility. The model captures the feedback in \Cref{fig:loop}: ownership and listing determine shared capacity, while vehicle availability, pickup delays, and owners' access reliability shape the incentives to own, share, and rent. Second, we derive individuals' best responses and characterize the resulting population partitions, including the emergence of different types of owners. Third, we characterize market equilibria, establish conditions for active sharing and changes in ownership, and specify a system of equations for equilibrium computation. Finally, an illustrative numerical study examines four case studies covering the impact of market conditions, platform operators, revenue incentives, and municipal regulation on market outcomes. Together, these analyses explain how sharing affects ownership and how platform and municipal objectives can support or undermine the outcome of ownership reduction.

\textit{Organization:} The remainder of this paper is structured as follows. In \Cref{sec:review} we review related work and position our contribution. In \Cref{sec:p2p} we present the market model and the definition of market equilibrium. In \Cref{sec:equ} we characterize individual choices, population partitions, and equilibria. In \Cref{sec:Results} we present results of four case studies and analyze the insights they provide in understanding the outcomes of such a market.
Finally, in \Cref{sec:Conclusion} we draw our conclusions and discuss directions for future research.
\section{Literature Review}\label{sec:review}
The analysis of the \gls{acr:p2p} \gls{acr:av} market lies at the intersection of several related streams of literature regarding \gls{acr:av}-enabled sharing, which we review in turn. First, the \gls{acr:amod} control literature studies how autonomous fleets can be operated efficiently through optimized matching, rebalancing, pooling, and pricing. Second, the mixed-autonomy and flexible-supply literature considers \gls{acr:mod} systems with both \glspl{acr:av} and human drivers. Human drivers operate in a decentralized manner and must be incentivized. Third, a growing literature on private \gls{acr:av} sharing and crowdsourcing examines the use of privately owned autonomous vehicles as a source of on-demand service capacity. Finally, the broader \gls{acr:p2p} sharing literature studies how access to a rental market affects consumers' decisions to own and share durable goods. 

\subsection{Control of \gls{acr:amod}} \label{sec:sub_control}
The development and advancement of \gls{acr:av} technology has motivated extensive research on \gls{acr:amod} service, where a fleet of autonomous vehicles provides on-demand passenger transportation. Compared to canonical human-driven \gls{acr:mod} (e.g., Uber and Lyft), the autonomous fleet enables centralized control without uncertainty from drivers' choice and behavior, and can thus improve the fleet control efficiency. A central topic in \gls{acr:amod}, therefore, lies in designing an effective fleet control algorithm to achieve operational efficiency, which often involves operational cost, passenger waiting, and service profit. Key topics in \gls{acr:amod} control focus on optimizing operations to improve performance on these metrics, through the optimization of passenger-vehicle matching~\citep{sadeghi2022reinforcement,levin2017general}, idle vehicle rebalancing~\citep{pavone2012robotic,huang2024bi,tresca2026robo}, ridepooling with passenger shared rides~\citep{alonso2017demand,li2026accelerating}, and spatial pricing~\citep{li2025learning,turan2020dynamic}. Recent studies also explore interaction between \gls{acr:amod} and other systems~\citep{li2026implementation,he2026hierarchical}. We refer interested readers to the review papers \citep{zardini2022analysis,li2026reproducibility} for a comprehensive introduction on the models and algorithms for \gls{acr:amod} control. These studies start from a fleet that is available to the operator or chosen as part of the system design, so that capacity is a dedicated resource to be allocated efficiently. In human-driven \gls{acr:mod}, by contrast, capacity is supplied by independent drivers who must first be attracted to the platform.


\subsection{Mixed autonomy in \gls{acr:mod}}
A second line of research studies the transition period, in which a \gls{acr:mod} fleet mixes \glspl{acr:av} and human drivers, as in the partnership between Uber and Waymo~\citep{waymo2024uber}. In a common formulation, the \gls{acr:av} fleet represents a committed or centrally controlled source of capacity, whereas human drivers are decentralized agents whose participation or operational decisions respond to economic incentives. The human drivers' choice to join is often modeled by comparing the expected earnings from joining the service platform with an external reservation payment. An equilibrium characterizing the number of active human drivers can then be derived from the platform's pricing, payment, and dispatch strategy between the \glspl{acr:av} and human drivers.

Following this paradigm, most of the works in this direction focus on analyzing the optimal strategy of the platform in controlling such mixed systems, or comparing the outcomes of the mixed system with different market configurations. For example, \citet{castro2024getting} build an equilibrium-constrained platform utility maximization model assuming an \gls{acr:av}-prioritized dispatching. The platform chooses an \gls{acr:av} size and a human driver wage to maximize its profit. In a follow-up work, they solve the platform profit maximization problem by designing an optimal dispatching strategy among the mixed fleet~\citep{castro2024autonomous}. Unlike these platform-oriented studies, \citet{freund2026supply} analyze market outcomes under four possible market configurations: platform-owned \glspl{acr:av} fleet with human driver, only platform-owned \glspl{acr:av}, leased \glspl{acr:av} with human driver, and leased \glspl{acr:av} and contracted human driver. Similarly, \citet{lian2025capturing} present and compare four possible market configurations, but characterized by different dispatching strategies and the number of \gls{acr:av} suppliers. The distinction between the committed capacity (\glspl{acr:av}) and flexible capacity (human drivers) is also closely related to the broader problem of mixed staffing in the on-demand economy, where flexible contractors provide capacity only when participation is sufficiently attractive, whereas a core workforce provides pre-committed capacity. The literature there also covers topics including the comparison of different staff compositions~\citep{lobel2024frontiers,dong2020managing} and optimal design of worker size and compensation, and price of service~\citep{gurvich2019operations,chakravarty2021blending}.

The mixed-autonomy and mixed-staffing perspectives both highlight an important departure from canonical \gls{acr:amod}: not all service capacity must be owned or directly controlled by the platform. With the advancement of autonomous driving, a private \gls{acr:av} can create a form of flexible supply when its owner lists idle capacity on the platform. This makes privately owned \glspl{acr:av} a potential source of shared mobility capacity, which we refer to as \gls{acr:p2p} sharing.

\subsection{Private \gls{acr:av} sharing}
\gls{acr:p2p} sharing market is one where individual owners list their idle capacity on a platform for renters to rent and earn money. 
This business model already exists in the carsharing market, for example on Turo in the United States and GoMore in Europe~\citep{dill2019effects,schwieterman2020estimating}. Since the advancement of autonomous driving technology and shared mobility services, some researchers have envisioned a potential \gls{acr:p2p} \gls{acr:av} business model where individual owners list their private \glspl{acr:av}~\citep{stocker2017shared}. While canonical carsharing is usually contracted on long-term rent from hours to days, \gls{acr:av} enables a more flexible trip-based sharing scheme, and coupled with on-demand service, this enables the \gls{acr:amod} service with private \glspl{acr:av}. 

Research on \gls{acr:p2p} private \gls{acr:av} sharing covers a wide range of topics from high-level market analysis to the development of service operation algorithms. \citet{zhang2025park} formulate the system equilibrium that includes both the \gls{acr:av} owners' choice between leaving their vehicle idle or sharing it, and the supply level dependent on the owner's choice. The equilibrium model is then used to characterize the profit maximization problem from the platform's perspective to decide the optimal price, rent payment, and platform-owned fleet size. Other works developed richer representations of the private-AV supply mechanism. For example, \citet{wang2021aggregate} develop an aggregate equilibrium model to analyze the \gls{acr:p2p} \gls{acr:av} sharing system, with extra consideration on the features of the service, including traveler's choice among multiple alternatives and the congestion effect of the service. \citet{siddiq2022ride} study the impact of individual-owned \glspl{acr:av} in a \gls{acr:mod} system on the platform's profit and the social welfare. They compare cases where the platform owns \glspl{acr:av} and the individual owns \glspl{acr:av}, and show the difference in their structural effect. While these works focus on aggregate market equilibrium and the impact of market conditions on the equilibrium, a recent study focuses on the operational aspect of the service~\citep{wang2026sharing}. They develop a time-expanded network flow model that captures the spatiotemporal heterogeneity in \gls{acr:av} usage of the service. This work complements the previous studies by characterizing the role of operation strategy in shaping the service performance.

Taken together, this literature establishes private \glspl{acr:av} as a plausible decentralized source of capacity for future on-demand platforms. It also studies increasingly detailed owner-sharing decisions and platform operations. A common feature of these models is that the population of private \gls{acr:av} owners is already present. The principal endogenous margin is whether and how an existing private \gls{acr:av} is shared. Since private \gls{acr:av} ownership is itself a decision, another question one can ask is how an individual chooses to acquire such a vehicle in the first place, considering that the sharing option exists. This distinction becomes important when expected sharing income is sufficiently large to alter the economics of vehicle ownership itself.

\subsection{\gls{acr:p2p} sharing and endogenous ownership}
The possibility that access to a sharing market changes product ownership has been studied in the broader literature on \gls{acr:p2p} product sharing. These studies explicitly consider the feedback between rental opportunities and the decision to own a durable good. \citet{fraiberger2017peer} are among the first to study the problem of ownership structure in a \gls{acr:p2p} sharing market for durable goods. They specifically accounted for the existence of a secondary market to sell the products, and they formally defined and characterized the market equilibrium in a \gls{acr:p2p} market with endogenous ownership. However, no analytical results exist to calculate such an equilibrium. \citet{jiang2018collaborative} work on a similar setup but ignored the matching friction. The demand is assumed to match the supply through a market-clearing price and no mismatch is considered. \citet{benjaafar2019peer} explicitly account for matching friction in their market equilibrium model, establish the uniqueness and existence of the equilibrium, and develop an analytical characterization of the ownership structure in the market. Their analysis shows that introducing \gls{acr:p2p} sharing can either increase or decrease aggregate ownership, depending on the ownership cost, service price, and rental payment. In the same spirit, \citet{filippas2020owning} distinguish between a short-run sharing equilibrium with fixed ownership and a long-run equilibrium in which consumers can adjust their ownership decisions. These studies provide the economic foundation for treating ownership and sharing as jointly determined decisions rather than taking the stock of shareable assets as fixed.

The general \gls{acr:p2p} product-sharing literature therefore captures an endogenous margin that is largely absent from the private \gls{acr:av} sharing literature. The sharing opportunity can change whether travelers choose to own the vehicle. At the same time, generic product-sharing models abstract from important characteristics of on-demand transportation. In an \gls{acr:amod} system with privately owned \glspl{acr:av}, the vehicle is not simply a durable product transferred from an owner to a renter for a predetermined rental period. Once supplied to a mobility platform, its availability to the owner depends on the state of the fleet and on ongoing passenger trips. Likewise, the value of the service to renters depends on the platform's ability to match demand with available vehicles.

This paper connects the gaps identified above. From the \gls{acr:p2p} product-sharing literature, we endogenize the vehicle-ownership decision together with the decision of an owner to release idle vehicle capacity. From the \gls{acr:amod} with private \gls{acr:av} sharing literature, we incorporate transportation-specific operational frictions into these decisions. On the supply side, an owner faces a reliability cost because a shared vehicle may not return in time when recalled, and this cost depends on fleet utilization and on trip durations. On the demand side, nonowners choose between the platform and an outside option, and the utilization of supply determines their waiting time. The resulting equilibrium determines who owns a private \gls{acr:av}, how much idle capacity is shared, and how much of it is used. In this sense, the paper sits between the transportation literature on privately supplied autonomous mobility and the \gls{acr:p2p} sharing literature on endogenous durable-good ownership.






\section{Model of the \gls{acr:av} sharing market} \label{sec:p2p}
In this section, we present our model of a \gls{acr:p2p} market for private \glspl{acr:av}.
We consider a population of~$M$ potential market participants distributed over a service area of size~$A$.
We represent the population as a continuum of unit mass, so that all aggregate quantities below are per capita. Individuals differ in their usage level, or \emph{type},~$\xi\sim F(\cdot)$, where~$F$ is a continuous distribution with full support on $[0,1]$. The type is the fraction of time during which the individual requires mobility, measured in vehicle time, and represents individuals' mobility demand. Following the \gls{acr:p2p} literature~\citep{benjaafar2019peer,filippas2020owning,fraiberger2017peer}, utility is linear in served usage, so that an individual of type~$\xi$ whose demand is fully served obtains utility~$\xi$. \cref{tab:nomenclature} summarizes the notation.

\begin{table}[t]
\centering
\caption{{Nomenclature.}}
\label{tab:nomenclature}
\small
\begin{tabular}{@{}ll@{\qquad}ll@{}}
\toprule
\multicolumn{2}{@{}l}{\textit{Population and preferences}} & \multicolumn{2}{l@{}}{\textit{Operations}}\\
$\xi$, $F$ & usage type, type distribution & $r$, $r_A$, $r_e$ & dispatch, acceptance, effective radius\\
$M$, $A$ & population size, service area & $v$, $m$ & vehicle speed, mean trip duration\\
$c$ & amortized ownership cost & $n_e$, $\bar w_e$ & population within $r_e$, maximum pickup time\\
$p$, $\gamma$ & fare per unit usage, commission & $I$ & idle vehicle stock per capita\\
$d_o$ & owner cost per unit committed time & $\beta$, $w$, $t$ & matching probability, pickup time, $w/m$\\
$d_r$ & renter inconvenience per unit usage & $S$, $V$ & listed supply, nonowner need\\
$v_{\mathrm{ext}}$ & per-unit utility of external option & $X$, $D$ & served passenger time, served trip rate\\
$v_w$ & waiting cost per unit time & $u_p$, $u_b$ & paid and committed utilization\\
$\mu$, $h$ & access-failure weight, recall tolerance & $\Delta_0$, $\delta$ & sharing surplus with and without waiting\\
\midrule
\multicolumn{2}{@{}l}{\textit{Individual choices}} & \multicolumn{2}{l@{}}{\textit{Equilibrium objects}}\\
$s$ & listed fraction of time & $a$, $b$ & net sharing income, access-penalty coefficient\\
$\mathcal O$, $O$ & ownership set, ownership share & $z$ & listing cutoff type\\
$R$ & nonowner payoff per unit usage & $\theta$, $\theta_L$, $\theta_H$, $\theta_0$ & ownership thresholds (active, no trade)\\
$G$ & ownership gain & & \\
\bottomrule
\end{tabular}
\end{table}

\paragraph{The sharing platform.} 
A platform intermediates the sharing of private \glspl{acr:av}. An owner of type~$\xi$ lists their vehicle for a fraction~$0\leq s\leq1-\xi$ of the time, and we call an owner who lists a \emph{sharer}. A nonowner either requests a shared vehicle, in which case we call them a \emph{renter}, or uses an external alternative. The platform charges renters a price~$p$ per unit of usage and retains a commission~$\gamma\in[0,1)$ of the fare, so a sharer earns~$(1-\gamma)p$ per unit of shared usage. As in canonical \gls{acr:amod}, the platform matches listed vehicles to requests, and it also manages the return of a vehicle when its owner recalls it. Unlike canonical \gls{acr:amod}, in which the platform owns a fleet of fixed size, supply depends on the owners' willingness to list. In the rest of this section, we describe the matching process and its frictions, the fleet accounting, and the owners' access to their vehicles, and we then derive the payoffs and define the market equilibrium.

\subsection{Service operation}\label{subsec:operation}
\paragraph{Matching strategy.}
A vehicle listed on the platform can be in one of three states: idle, en-route, and on-board. An idle vehicle has not been matched to any passenger; a vehicle that is en-route is matched with a passenger and on its way to pick them up; and an on-board vehicle has already picked up the matched passenger and is transporting them to their destination. Only idle vehicles can be matched to passenger requests. En-route and on-board vehicles have both committed their capacity to certain passenger services. The platform applies a maximum radius matching policy to pair passengers with the nearest idle vehicle. When a passenger sends a request on the platform, the platform first checks for any idle vehicles within a specified dispatch radius $r>0$. If an idle vehicle exists, the platform dispatches the closest one to the requesting passenger. 
The radius~$r$ is a policy parameter of the platform, which we take as exogenous.
Maximum-radius matching is a common policy in spatial matching models of on-demand services~\citep{yang2020optimizing,li2024aggregate}.

\paragraph{Nonowner choice.} When a nonowner needs transportation, they can either use the sharing service or an external option with per-unit utility $v_{\mathrm{ext}}$. 
For example,~$v_{\mathrm{ext}}$ reflects the quality of other modes such as transit and micromobility.
A nonowner makes a choice between these two options by comparing the payoff through the following process.
First, they check the sharing platform for the availability of an idle vehicle. 
If they find one, they assess the utility of using the sharing service based on the price and the waiting time until pickup. Specifically, when using the sharing service, a nonowner pays $p$ per unit of usage of the shared vehicles, and bears an inconvenience cost $d_r \in [0,1)$ plus the cost of waiting, valued at $v_w$ per unit time. The inconvenience cost $d_r$ measures the disutility of using others' vehicles compared to their own vehicle. They then compare this utility with that of the external option and choose the one with higher utility. If they cannot find an available vehicle on the platform, they use the external option.

To retain tractability, we assume that all vehicles travel at a fixed average speed $v>0$, and the passenger trip has a mean duration of $m$. With fixed speed $v$, a vehicle at distance $L$ away from the passenger has quoted pickup time $P = L/v$, so a nonowner chooses sharing service if 
\begin{equation}\label{eq:accept}
 1-p-d_r-\frac{v_wP}{m}\geq v_{\mathrm{ext}}
 \quad\Longleftrightarrow\quad
 P\leq\frac{m(1-p-d_r-v_{\mathrm{ext}})}{v_w},
\end{equation}
where $1-p-d_r-\frac{v_wP}{m}$\footnote{Dividing $P$ by the mean trip duration~$m$ converts the waiting cost of a trip into a cost per unit of usage.} is the unit utility of using the sharing service conditioned on successful match.
Write~$\Delta_0:=1-p-d_r-v_{\mathrm{ext}}$. We make the following assumption on admissible parameter values.
\begin{assumption}[Admissibility]\label{assump:admi}
    $p+d_r+v_{\mathrm{ext}}<1$.
\end{assumption}
\cref{assump:admi} states that the sharing service must improve on the external option before pickup waiting is added. If it does not hold, no positive pickup time is acceptable to the nonowners, and it is always preferable for them to use the external option. In this case, the sharing market does not exist.

Substituting~$P=L/v$ into \eqref{eq:accept} yields the acceptance radius
\begin{equation}
    L \leq r_A, \quad r_A := \frac{vm\Delta_0}{v_w}.
\end{equation}
Passengers reject pickups from beyond~$r_A$, so increasing the dispatch radius~$r$ beyond~$r_A$ does not increase the number of matches.
We can therefore define an effective service radius by $r_e = \min\{r,r_A\}$, which caps the matching radius at the passenger acceptable value. 
The population within the effective radius is
\begin{equation}\label{eq:idle}
    n_e = \pi r_e^2\frac{M}{A},
\end{equation}
and the maximum waiting time of a matched passenger is $\overline{w}_e=\frac{r_e}{v}$.

\paragraph{Matching probability and expected waiting.} Not every request can be matched to an idle vehicle, and the matching probability depends on the idle vehicle stock. Let $I$ denote the mean idle vehicle stock measured in vehicle time per capita. Following spatial Poisson approximations used in on-demand
matching models \citep{zhang2021pool,li2024aggregate}, we
approximate the locations of idle vehicles seen by a passenger through a homogeneous Poisson process with intensity $\frac{M}{A}I$. This model is a stationary spatial approximation that captures the number of idle vehicles within a certain distance from the passenger by modeling it as a counting process. With $n_eI$ representing the idle vehicle stock within the effective matching radius, the probability of a successful match, which is the probability of at least one idle vehicle within the matching radius, and the expected pickup time conditional on successful match can be calculated by
\begin{align}
 \beta(I)&=1-e^{-n_eI},\label{eq:beta}\\
 w(I)&=\bar w_e\,
 \frac{\displaystyle\int_0^1 e^{-n_eIz^2}\,dz-e^{-n_eI}}
 {1-e^{-n_eI}},\qquad I>0.\label{eq:pickup}
\end{align}
Both values depend on idle vehicle stock $I$.
When no confusion arises, we drop the argument~$I$.
The derivation of~$\beta$ and~$w$ is provided in \cref{app:matching}, where \cref{lem:pickup} shows that matching probability~$\beta$ increases with idle stock and the expected pickup time~$w$ decreases.


\subsection{Fleet accounting}
\Cref{subsec:operation} introduces the three states of a listed vehicle: idle, en-route, and on-board. There is a conservation relationship between the total shared vehicle time and the vehicle time spent in the three states. To see this, let $S$ denote the aggregate listed vehicle time per capita and $V$ denote aggregate transportation need of all nonowners. Only the successfully matched and accepted portion is offered the sharing service, so not all transportation needs are satisfied by the sharing service. Denote the accepted need by $X=\beta V$. The external option accounts for the rest $(1-\beta)V$. Denote the served trip rate as $D=X/m$. Fleet time conservation gives
\begin{equation}\label{eq:conservation}
    S = I+Dw+Dm=I+(1+t)X, \quad t=\frac{w}{m}.
\end{equation}
The three terms on the right-hand side of the first equality in \eqref{eq:conservation} are idle vehicle time, pickup time, and passenger-on-board time. This separation is important, since only passenger-on-board time generates revenue, while both pickup and passenger-on-board time induce operation cost. For $S>0$, we define paid and committed utilization of shared vehicle capacity by
\begin{align}
 u_p=\frac{Dm}{S}&=\frac XS,\label{eq:paid}\\
 u_b=\frac{D(m+w)}S&=(1+t)u_p=1-\frac IS.\label{eq:committed}
\end{align}
The pickup fraction is then $u_b-u_p$. 
When~$S=0$, no capacity is listed and we set~$u_p=u_b=0$ by convention. 
The pair~$(I,u_p)$ fully characterizes the fleet accounting and we refer to it as the \emph{operation state}.
\begin{remark}[Elastic activation of shared capacity]
In \eqref{eq:conservation}, operation cost accumulates only on pickup and passenger-on-board time, not on the idle time $I$. This departs from canonical \gls{acr:amod} operation models, where fleet size is pre-committed, and every deployed vehicle is assumed to cruise for the full operating period, so cost is charged against fleet-hours regardless of realized demand \citep{zhang2016control}. Under \gls{acr:p2p} sharing, supply is contributed in reversible increments and need not be fully activated. While idle vehicles can still be rebalanced and create operational costs, rebalancing concerns the real-time spatiotemporal demand distribution, so it is not considered here in the aggregate stationary model.
\end{remark}

\subsection{Owner access}
The willingness of owners to share depents on recall reliability.
When owners list their vehicle, they are exposed to uncertainty about access when an unplanned need arises.
This feature is specific to \gls{acr:p2p} vehicle sharing, and it is usually absent from general \gls{acr:p2p} sharing studies.
We model this concern through an access check performed at a time drawn independently of the state of the vehicle.
When an owner wants their vehicle back, they check whether their vehicle is idle, and if not, whether the current service commitment can finish within a tolerance $h \in [0,m]$. We assume that there is a fixed allowance for the vehicle to return to the owner after it becomes idle. 
This allowance captures the expected return time from service region to owner's location and is agreed upon when the owner lists their vehicle on the platform. For the owner's recall check, an idle vehicle has no outstanding service commitment and passes the check. A vehicle undertaking pickup or passenger service passes if its remaining commitment does not exceed $h$.

\begin{figure}[tb]
\centering
\begin{tikzpicture}[x=1.15cm,y=1cm,font=\small]
 \fill[gray!8] (0,0) rectangle (1.6,.7);
 \fill[gray!32] (1.6,0) rectangle (3.8,.7);
 \fill[gray!32] (3.8,0) rectangle (7,.7);
 \fill[gray!8] (7,0) rectangle (8.8,.7);
 \draw (0,0) rectangle (8.8,.7);
 \foreach \a in {1.6,3.8,7} {\draw (\a,0)--(\a,.7);}
 \node at (.8,.35) {Idle};
 \node at (2.7,.35) {Pickup $P$};
 \node at (5.4,.35) {Passenger $m-h$};
 \node at (7.9,.35) {Passenger $h$};
 \draw[<->] (1.6,-.25)--(7,-.25);
 \node at (4.3,-.57) {Recall failure: $P+m-h$};
\end{tikzpicture}
\caption{A service cycle and its recall-failure portion. The last
$h$ units of passenger time are within the owner's completion tolerance.}
\label{fig:timeline}
\end{figure}
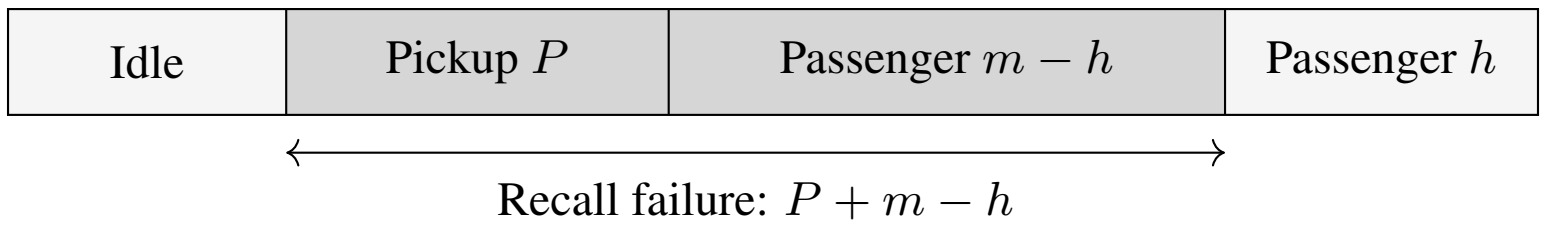

\Cref{fig:timeline} is a schematic presentation of the service cycle and its recall-failure portion. With trip duration $m$, the pickup and the first $m-h$ of passenger service fail the recall check. Therefore, an owner listing fraction $s$ has access reliability
\begin{equation} \label{eq:reliability}
     \rho(s)=1-s\left(u_b-\frac hm u_p\right)
        =1-su_p(1-h/m+t).
\end{equation}
The sharing fraction $s$ accounts for the proportion of vehicle time exposed to platform service. 
Within this duration, the vehicle has service commitment for $u_b$ of the time, and $\frac{h}{m}u_p$ of the time passes the recall check. 
Therefore, $u_b - \frac{h}{m}u_p$ is the fraction of the listed vehicle time that does not pass an owner recall check, and~$s(u_b - \frac{h}{m}u_p)$ is the probability that a recall fails.
The recall failure probability enters the owner's payoff, which we derive in the next section. Note that the access check is independent of the demand on the platform.
If owner's own needs tend to coincide with demand peaks, this assumption overestimates reliability.

\begin{remark}[General trip-duration distributions]\label{rem:tripdist}
Expression \eqref{eq:reliability} uses the fact that, with trips of fixed duration $m$, a fraction $1-h/m$ of on-board time fails the recall check. If trip durations are random with distribution $T$ and mean $m=\mathbb E[T]$, the remaining on-board time at a random time during a trip follows the equilibrium-excess distribution of $T$, and the fraction of on-board time that fails the check becomes
\[
\kappa_h\coloneqq\frac{\mathbb E[(T-h)^+]}{\mathbb E[T]}\in[0,1],
\]
with $\kappa_h=(1-h/m)^+$ for deterministic trips and $\kappa_h=e^{-h/m}$ for exponential trips. Keeping the convention that a vehicle in pickup fails the check, all results of this paper continue to hold with $1-h/m$ replaced by $\kappa_h$, because they only use $0\leq\kappa_h\leq1$ and $\kappa_h+t>0$. 

\end{remark}

\subsection{Payoffs}
\paragraph{Owner payoff.} When an owner lists, they get a payment of $(1-\gamma)p$ per unit of shared usage of their vehicle. They also bear a wear-and-tear cost $d_o \in [0,1)$ per unit of committed time, and disutility for recall failure during the sharing interval. The payoff for an owner of type $\xi$ who shares $s$ is then
\begin{align}
 & \pi_o(\xi;s) \nonumber \\
 &=\xi+s\{(1-\gamma)p u_p-d_o u_b
                  -\mu\xi[u_b-(h/m)u_p]\}-c \label{eq:ownerpayoff_exp} \\
 &=\xi+su_p[a-b\xi]-c,\label{eq:ownerpayoff}
\end{align}
with $a:=(1-\gamma)p-d_o(1+t)$ and $b:=\mu(1-h/m+t)$. The first term in \eqref{eq:ownerpayoff_exp} is the utility of a sharer's own usage of their private vehicle. They use their vehicle to meet their mobility demand $\xi$ and derive a utility of $\xi$ from it. The second term is the expected utility from sharing. Within the expression, $(1-\gamma)p u_p$ is the expected rental income per listed capacity, $d_o u_b$ the cost of service commitment, and $\mu\xi[u_b-(h/m)u_p]$ the reliability loss due to the failure of recall. We model the loss an owner places on lost access to their vehicle as \(\mu\xi\), increasing proportionally with their own-use type \(\xi\). 
The last term is the amortized private \gls{acr:av} ownership cost, which includes the purchase price,maintenance, and insurance. 
After reordering the terms in \eqref{eq:ownerpayoff},~$a$ is the net sharing income per unit of paid utilization, and~$b\xi$ is the access penalty per unit of paid utilization. 
An owner decides a sharing fraction $s$ that maximizes their payoff \eqref{eq:ownerpayoff}.

\paragraph{Nonowner payoff.} A nonowner first checks availability on the sharing platform when transportation is needed. According to the discussion in \Cref{subsec:operation}, under effective service radius $r_e$, using the sharing option is always no worse than the external option. Therefore, if a nonowner can be successfully matched to an idle vehicle, they always choose to use sharing service, and if not, they use the external option. Denote the per unit gap between the utility of using sharing and external option by $\delta := \Delta_0-v_wt=1-p-d_r-v_{\mathrm{ext}}-v_wt$. The payoff of a nonowner is
\begin{equation}\label{eq:nonownerp}
    \pi_r(\xi) = \xi R, \quad R=v_{\mathrm{ext}}+\beta\delta.
\end{equation}
The expected benefit of using sharing over relying entirely on the external option is $\xi\beta\delta$.

\subsection{Market equilibrium}
As discussed in the previous sections, the service operations in a \gls{acr:p2p} \gls{acr:av} sharing market depend on how many people own vehicles, how much vehicle time owners list, and how much of the nonowners' transportation need is met through the platform.
These decisions, in turn, depend on the service conditions they collectively generate. Vehicle availability and pickup time determine the attractiveness of traveling without ownership, while fleet utilization and service commitment determine owners' sharing income and access reliability. A stationary market equilibrium requires consistency between these individual decisions and aggregate operations. 
Individuals choose ownership, listing, and platform use optimally given prevailing service conditions, and their choices generate the supply and demand that sustain those same conditions. 
We formalize this relationship below, after stating our assumption on the primitives.

\begin{assumption}[Primitives]\label{assump:primitives}
The parameters satisfy $M,A,r,v,m,v_w>0$, $c>0$, $p>0$, $\mu>0$, $h\in[0,m]$, and $\gamma,d_o,d_r,v_{\mathrm{ext}}\in[0,1)$.
\end{assumption}
\Cref{assump:primitives} is not restrictive. It defines the natural range of the parameters in a real-word system. For example, the condition $c>0$ states that owning a vehicle is costly, and the condition $\mu>0$ states that owners value access to their vehicle.

\begin{definition}[Market equilibrium]\label{def:equilibrium}
Fix a market represented by 
\[\mathcal M=
 (M,A,F,c,p,\gamma,d_o,d_r,v_{\mathrm{ext}},v_w,m,\mu,h,r,v).\]
 Here, $M$, $A$, and $F$ describe the population size, service area, and distribution of personal transportation needs. $c$ is the amortized ownership cost. $p$ and $\gamma$ specify the fare and platform commission. $d_o$ is the incremental operating cost of shared driving. Passenger preferences are described by the inconvenience cost $d_r$ of using a shared vehicle, the external option utility $v_\mathrm{ext}$, and waiting cost $v_w$. The remaining parameters specify the mean passenger trip duration $m$, owners' recall delay tolerance $h$ and the access failure disutility weight $\mu$, the platform dispatch radius $r$, and vehicle driving speed $v$. 

 A market equilibrium consists of a measurable ownership set $\mathcal O\subseteq[0,1]$, listing proportion $s(\xi) \in [0,1-\xi]$ for owners, and nonnegative aggregates~$(S,V,I,X,D,u_p,u_b)$.
 Here,~$S$ is the listed vehicle time per capita, $V$ is the total nonowner transportation use per capita, and $I$ is the idle vehicle stock per capita. 
 The quantities $X$ and $D=X/m$ measure served passenger time and the served trip rate, and $u_p$ and $u_b$ are the paid and committed fractions of listed vehicle time. A market equilibrium of a given market $\mathcal M$ is defined by the following conditions.
 \begin{enumerate}
     \item \emph{Optimal ownership and listing.} Ownership, nonownership, and listing decisions must be optimal for the individuals making them. Concretely, define the optimized ownership payoff for a type $\xi$ individual by
     \[\pi_o^*(\xi) =\max_{0\leq s\leq1-\xi} \left\{\xi-c+s\left[(1-\gamma)p\,u_p-d_o u_b -\mu\xi\left(u_b-\frac hm u_p\right)\right]\right\}.\]
     It must hold that
     \[
     \begin{aligned}
     \pi_o^*(\xi)\geq\xi R(I),&&\xi\in\mathcal O,\quad
     \pi_o^*(\xi)<\xi R(I),&&\xi\notin\mathcal O,
     \end{aligned}
    \]
    and for $\xi \in \mathcal{O}$
    \[ s(\xi)\in\operatorname*{arg\,max}_{0\leq s\leq1-\xi}
 \pi_o(\xi;s).\]
    \item \emph{Aggregation of individual choices.} The listed supply and potential passenger demand equal the quantities generated by the ownership and listing decisions:
    \[ S=\int_{\mathcal O}s(\xi)\,dF(\xi),\qquad
       V=\int_{[0,1]\setminus\mathcal O}\xi\,dF(\xi).\]
The passenger time served by the platform satisfies~$X=\beta V$,~$D=X/m.$
    \item \emph{Fleet and utilization consistency.} Listed vehicle time is equal to the sum of idle vehicle time, pickup time, and passenger on-board time~$S=I+Dw+Dm.$
    Paid and committed utilization must therefore satisfy (with~$u_p=u_b=0$ if~$S=0$)
    \[ u_p=\frac{Dm}{S},\qquad
       u_b=\frac{D[m+w]}{S}=1-\frac IS.\]
 \end{enumerate}
\end{definition}
The equilibrium definition closes the loop of \cref{fig:loop}. Ownership and listing determine available capacity and nonowner transportation needs. Fleet use determines matching and passenger acceptance, which then determine the earnings and access reliability that support the original ownership and sharing choices. 
\section{Market equilibrium analysis} \label{sec:equ}
In this section, we characterize the market equilibria of \Cref{sec:p2p}. We first derive individual choices and the population partitions they induce under given service conditions. We then impose aggregation and fleet consistency to identify possible equilibria. Finally, we analyze the ownership and operation implications of the different equilibria.

\subsection{Optimal listing of the owner}\label{subsec:listing}
At equilibrium, an owner of type $\xi$ lists capacity $s$ that maximizes their expected payoff in \eqref{eq:ownerpayoff}. The next proposition shows that, at a given operation state~$(I,u_p)$, an owner lists either all or none of their idle time.
\begin{proposition}[Optimal listing]\label{prop:listing}
Given $I$ and $u_p>0$, the optimal listed capacity for an owner of type $\xi$ is
\begin{equation}\label{eq:listchoice}
 s^*(\xi)=
 \begin{cases}
  1-\xi,&a-b\xi>0,\\
  0,&a-b\xi<0.
 \end{cases}
\end{equation}
At equality, every~$s\in [0,1-\xi]$ is optimal.
If $a>0$, the willingness to list cutoff is
\begin{equation}\label{eq:cutoff}
    z = \min\{1,a/b\}.
\end{equation}
If $a\leq 0$, no positive mass of owners strictly prefers listing.
\end{proposition}
\begin{proof}
For given $I$ and $u_p$, the owner payoff \eqref{eq:ownerpayoff} is affine in $s\in[0,1-\xi]$, with slope $u_p[a-b\xi]$. A positive slope $a-b\xi>0$ makes full listing $s=1-\xi$ optimal, a negative slope $a-b\xi<0$ makes withholding $s=0$ optimal, and zero slope makes every feasible listing level optimal. 
\end{proof}
In words, an owner lists all idle time when the access penalty per unit of paid utilization,~$b\xi$, is smaller than the net sharing income~$a$, and withholds otherwise.
Since~$a$ decreases and~$b$ increases with the pickup ratio~$t$, longer pickups lower the cutoff \eqref{eq:cutoff}, so fewer owners list.

\subsection{Population partition} \label{sec:population}
In a \gls{acr:p2p} \gls{acr:av} sharing market, the individuals in the population choose to be an \gls{acr:av} owner or not. They also decide, if they are an owner, how much capacity to list on the sharing platform; if they do not own an \gls{acr:av}, whether they rent through the platform or use the external option to meet their transportation needs. 
These choices partition the population into four segments: owners who share, owners who withhold, renters, and nonowners who use the external option. Since supply and rental demand follow from this partition by aggregation, we characterize it first.

\begin{figure}[tb]
\centering
\input{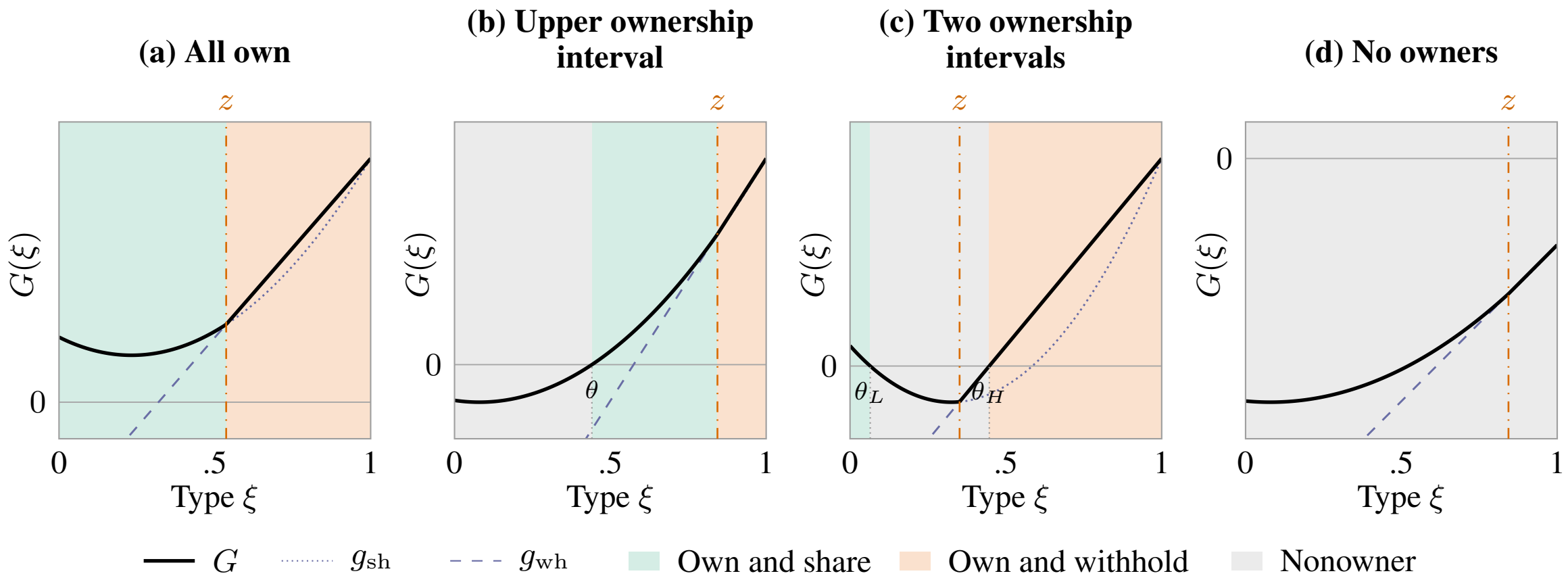}
\caption{Population choices at a given {operation} state $(I,u_p)${, for the four cases of \Cref{theorem:partition}}. The solid
curve is {the ownership gain} $G=\max\{g_{\mathrm{sh}},g_{\mathrm{wh}}\}${ in \eqref{eq:ownergain}, the dotted and dashed curves are its sharing and withholding branches, and the dashed line marks the listing cutoff $z$. Shading shows the resulting roles.}}
\label{fig:partition}
\end{figure}

The ownership decision is governed by the payoff gap between owning and not owning. For a given operation state $(I,u_p)$, define the payoff gain of owning a vehicle compared to not owning by
\begin{align}
 g_{\mathrm{sh}}(\xi)
 &=u_pb\xi^2+[1-R-u_p(a+b)]\xi+u_pa-c,\nonumber\\
 g_{\mathrm{wh}}(\xi)&=(1-R)\xi-c,\label{eq:ownergain} \\
 G(\xi)&=\max\{g_{\mathrm{sh}},g_{\mathrm{wh}}\}. \nonumber
\end{align}
The first two equations in \eqref{eq:ownergain} capture the sharing and withholding branches of the ownership gain, respectively. $G(\xi)$ is the final ownership gain: the sharing branch $g_{sh}$ applies for individual type $\xi$ below sharing cutoff $z$, and the withholding branch above it. An individual with $G(\xi)>0$ prefers to be an owner, an individual with $G(\xi)<0$ prefers to be a nonowner, and indifferent types have zero mass.
The following theorem gives the possible partitions at a given operation state.
\begin{theorem}[Population partition]\label{theorem:partition}
Given a feasible operation state~$(I,u_p)$ with $I>0$ and~$0<u_p<1/[1+t(I)]$, exactly one of the following ownership partitions occurs:
\begin{enumerate}
    \item All individuals own, if $\min G \geq 0$.
    \item Owners form one upper interval $[\theta,1]$, if $G(0) \leq 0 < G(1)$ and $\min G<0$.
    \item Owners form two intervals $[0,\theta_L] \cup [\theta_H,1]$, if $G(0)>0$ and $\min G<0$.
    \item No individuals own, if $G(1) \leq 0$.
\end{enumerate}
\Cref{fig:partition} illustrates the four configurations. Within the ownership set, types below cutoff $z$ list their full available capacity and types above $z$ withhold (\Cref{prop:listing}). Specifically, in the two interval case, $0<\theta_L<\theta_H<1$ and every lower interval owner shares. 
\end{theorem}

The proof is provided in Appendix~\ref{app:partition}, and it relies on the convexity of $G$ and on the ordering $G(0)<G(1)$ of its endpoints. Beyond ownership, \Cref{fig:partition} shows which owners share. With upper-interval ownership (\Cref{fig:partition}(b)), nonowners have types below $\theta$, owners in $(\theta,z)$ share if $\theta<z$, and owners above $\max\{\theta,z\}$ withhold. If $\theta\geq z$, every owner withholds. With two-interval ownership (\Cref{fig:partition}(c)), the low-use owners in $[0,\theta_L]$ share, types in $(\theta_L,\theta_H)$ do not own, and the high-use owners withhold if $\theta_H\geq z$. If instead $\theta_H<z$, the high-use owners in $(\theta_H,z)$ also share, before the withholding group. In all cases, listing follows the rule of \Cref{prop:listing}.

The four possible configurations in \Cref{theorem:partition} are derived based on a given service state $(I,u_p)$. A complete equilibrium analysis also requires that the partition can support the given service state. In the next subsection, we impose the aggregation and consistency conditions to characterize possible equilibria that can arise in the system.

\subsection{Equilibrium characterization and computation} \label{sec:characterization}
We start by classifying equilibria by the served trip rate~$D$.
An equilibrium is a \emph{no-trade equilibrium} if~$D=0$, i.e., no nonowner transportation need is served by the platform, and an \emph{active equilibrium} if~$D>0$.
The next lemma shows that listed supply and trade vanish together. 
\begin{lemma}[Listed supply in no-trade and active equilibria]\label{co:listed}
In every equilibrium,~$D>0$ implies~$S>0$ and~$D=0$ implies~$S=0$.
\end{lemma}
The proof is in \Cref{app:supply}. We next discuss the characterization and computation of the two types of equilibrium.

\subsubsection{No trade equilibrium} 

By \Cref{co:listed}, a no-trade equilibrium has $D=0$ and $S=0$, i.e., there is no listed supply on the platform and no transportation need is satisfied by the platform. This makes the no trade equilibrium equivalent to a market without the sharing platform. Individuals therefore compare the owner payoff $\xi-c$ with the external option payoff $\xi v_{\mathrm{ext}}$. The resulting ownership threshold, population profile, and ownership share are
\begin{equation}\label{eq:notrade}
 \theta_0=\frac{c}{1-v_{\mathrm{ext}}},\qquad
 \mathcal O_0=\{\xi\in[0,1]:\xi\geq\theta_0\},\qquad
 O_0=1-F(\min\{\theta_0,1\}).
\end{equation}
Both owners and nonowners exist if~$c<1-v_{\mathrm{ext}}$, and nobody owns otherwise. A no-trade equilibrium exists in every market and is unique among equilibria with $D=0$. Existence follows because, with $S=I=0$, no nonowner can be matched, so $R=v_{\mathrm{ext}}$, owners earn no sharing income, and the threshold rule \eqref{eq:notrade} with $s=0$ is a best response for every type. Uniqueness follows because the threshold $\theta_0$ is unique and all other aggregates in \Cref{def:equilibrium} are then determined by their definitions. The no-trade equilibrium is the coordination failure familiar from two-sided markets~\citep{caillaud2003chicken,evans2010failure}. No owner lists because no renter can be served, and no renter can be served because no owner lists. Whether an active equilibrium coexists with it is therefore a question of market launch, which we address at the end of this section. We use the no-trade equilibrium as the benchmark against which we assess active equilibria, so that differences measure the impact of the sharing platform.


\subsubsection{Active equilibrium} \label{subsec:active}
\paragraph{Characterization of an active equilibrium.} At an active equilibrium, a positive mass of supply exists ($S>0$) and some nonowners use the platform for their transportation need ($D>0$). This indicates that the ownership structure must be one of the upper interval or two interval configurations in \cref{fig:partition}(b)(c) of \cref{theorem:partition}. In \Cref{theorem:partition}, these two configurations are distinguished by the ownership gain at zero personal use,~$G(0)$. 
When $G(0)\leq 0$, an active equilibrium has upper interval structure, and it has two interval structure otherwise. This can be summarized as
\begin{align}
 \text{Upper-interval:}\quad
 &\mathcal O=[\theta,1],\quad 0<\theta<z, \quad
   u_pa\leq c;\label{eq:upper}\\
 \text{Two-interval:}\quad
 &\mathcal O=[0,\theta_L]\cup[\theta_H,1], \quad
   u_pa>c.\label{eq:twointervals}
\end{align}
\eqref{eq:upper} and \eqref{eq:twointervals} are classifications of possible active equilibria. They do not assert the existence of either structure based on market parameters, as $u_p$ and $a$ are endogenously decided. In the first case, nonowners, sharers, and withholders appear in increasing order of personal transportation need. In the second, types below $\theta_L$ own and share, types between $\theta_L$ and $\theta_H$ remain nonowners, and types above $\theta_H$ own. Within the upper ownership of a two interval structure, types below $z$ share and types above $z$ withhold. A feature of the equilibrium with two-interval ownership is the appearance of low-use owners. These owners buy vehicles primarily to earn money by sharing, so we refer to them as \emph{investment owners}.

Which type of ownership structure can arise at an active equilibrium depends on ownership costs and owners' access penalty.
\begin{proposition}[Sufficient conditions for upper interval ownership]\label{prop:upper_sufficient}
An active equilibrium has upper interval ownership if: 1. $c\geq (1-\gamma)p-d_o$, or 2. $\mu\leq\gamma p+d_o+d_r$.
\end{proposition}

The proof is presented in \cref{app:ownership}. Both conditions in \cref{prop:upper_sufficient} have a transparent interpretation. 
The first condition states that sharing income cannot cover the ownership cost even with full utilization and no pickups, so no type buys a vehicle only to share it. 
For the second condition, we call~$\gamma p+d_o+d_r$ the \emph{transaction wedge}. 
A unit of usage served through the platform costs the renter~$p+d_r$, while the owner who supplies it earns at most~$(1-\gamma)p-d_o$, so the wedge is the value lost when a unit of mobility is traded with the nonowners. The proof shows that the slope of $G$ on the sharing branch exceeds $\gamma p+d_o+d_r-\mu$. 
The wedge thus makes an additional unit of personal usage more valuable to an owner than to a nonowner, whereas the access penalty makes sharing increasingly costly as usage grows. 
If the penalty weight $\mu$ does not exceed the wedge, $G$ is strictly increasing, and owners always form an upper interval. 
Investment owners can therefore appear only if recall reliability is sufficiently valuable to owners, a friction that is specific to on-demand mobility.

\begin{corollary}[Necessary conditions for two interval ownership]\label{co:two_necessary}
An active equilibrium with two interval ownership must satisfy
\[c<(1-\gamma)p-d_o,\mu>\gamma p+d_o+d_r, \text{ and }t<\frac{(1-\gamma)p}{c+d_o}-1.\]
\end{corollary}
The first two inequalities require sufficiently low ownership costs and a sufficiently large access penalty for low-use owners to coexist with intermediate-use nonowners. 
The third inequality limits unpaid pickup commitments, and it ensures that the ownership cost is below the sharing earnings attainable with no idle time at the equilibrium pickup ratio. Details on the proof of \Cref{co:two_necessary} are presented in Appendix \ref{app:ownership}. 

Note that the two ownership structures can arise at different active equilibria of the same market.
Next, we compare individual payoffs associated with these two types of active equilibria to those in a no trade equilibrium, and we examine the ownership effects of each structure.

\subsubsection{Comparison between active and no trade equilibrium}
\begin{proposition}[Comparison of participant payoff]\label{prop:payoff}
At every active equilibrium, each individual's optimized payoff is weakly higher than at the no-trade equilibrium, and a positive population mass obtains a strictly higher payoff. This holds independent of the ownership structure type.
\end{proposition}
The proof is provided in Appendix \ref{app:payoff}. 
The intuition is that the platform only adds options.
Individuals who would not own without sharing strictly benefit from the improved access provided by the service.
Individuals who would own without sharing can list their vehicle when sharing is profitable, or withhold it and preserve their private use.
Note that \cref{prop:payoff} concerns private payoffs.
It does not account for externalities, e.g., the additional vehicle stock or the unoccupied pickup travel generated by the service~\citep{nair2020model,henao2019impact}, so it does not imply that an active equilibrium is socially preferable to no-trade.
We next examine the ownership change at an active equilibrium compared to no trade, and the conditions that determine this ownership change. We split the discussion based on the ownership structure of the active equilibrium.

\paragraph{Upper interval ownership.} At an active equilibrium with upper interval ownership, the marginal type $\theta$ is indifferent between owning and not owning. Denote the operation state at the active equilibrium as $(I,u_p)$. Since $\theta < z$, this individual would list all available vehicle time when owning. Equating the owner and nonowner payoff for this type gives~$\theta-c+u_p(1-\theta)(a-b\theta)=\theta R$.
Substituting $R=v_{\mathrm{ext}}+\beta\delta$ and $c=(1-v_{\mathrm{ext}})\theta_0$ and rearranging gives
\begin{equation}\label{eq:decomposition}
 (1-v_{\mathrm{ext}})(\theta_0-\theta)
 =u_p(1-\theta)(a-b\theta)-\theta\beta\delta.
\end{equation}
The first term on the right-hand side is the marginal type's net benefit from listing as an owner, and the second is the same type's expected benefit from platform access as a nonowner, relative to the external option. \cref{eq:decomposition} shows that if the listing benefit exceeds the access benefit, then~$\theta < \theta_0$ and ownership increases relative to no-trade.
Conversely, if the access benefit dominates, then~$\theta>\theta_0$ and ownership decreases.

The same comparison can be understood through the ownership incentive of the individual of type $\theta_0$ who is indifferent about ownership without the platform. The ownership share at the no-trade equilibrium is $O_0=1-F(\theta_0)$. With upper interval ownership, the ownership share is $O=1-F(\theta)$. Since $G$ is negative below $\theta$ and positive above it at operation state $(I,u_p)$, the sign of ownership change is
\[
 \operatorname{sgn}(O-O_0)
 =\operatorname{sgn}(\theta_0-\theta)
 =\operatorname{sgn}G(\theta_0;I,u_p).
\]
Thus the ownership comparison can be read from the ownership payoff gap $G$ of the no-trade marginal individual $\theta_0$, evaluated at the operation state $(I,u_p)$ at the active equilibrium.

\paragraph{Two interval ownership.}At an active equilibrium with two interval ownership, individuals at both low and high ownership margins are indifferent between owning and not owning. Equating the owner and nonowner payoff for both margins gives
\[\theta_j-c+u_p(1-\theta_j)[a-b\theta_j]^+=\theta_j R,\qquad j\in\{L,H\}.\]
$\theta_L<z$, so owners in the lower interval list. For owners in the upper interval, a useful special case is $\theta_H\geq z$, where all upper interval owners withhold. Their payoff condition reduces to $\theta_H-c=\theta_HR$, which gives
\[\theta_H=\frac{c}{1-R}>\frac{c}{1-v_{\mathrm{ext}}}=\theta_0.\]
In words, improved access without ownership displaces personal ownership at the upper margin, while low-use individuals acquire vehicles in order to share them.
In this case, all listed capacity is supplied by investment owners and every high-use owner withholds, so the shared fleet consists entirely of vehicles bought to be shared.
The total ownership change relative to no-trade is
\begin{equation}\label{sp:investdecomposition}
 O-O_0=\underbrace{F(\theta_0)-F(\theta_H)}_{\text{high-use owner change}}
       +\underbrace{F(\theta_L)}_{\text{investment owners}}.
\end{equation}
When $\theta_L <\theta_0<\theta_H$, the comparison is between the new investment owners of mass $F(\theta_L)$, and former owners who no longer own, of mass $F(\theta_H)-F(\theta_0)$. 
Investment ownership therefore need not increase total ownership, because improved access can lead more former owners to give up their vehicle.

\paragraph{Computation.} We now collect the conditions that jointly determine ownership, listed supply, and passenger service at an active equilibrium based on the previous discussion. The following conditions combine individual choices, population aggregation, and the operational state of the service that must all hold true at an active equilibrium $(\mathcal O,s,S,V,I,X,D,u_p,u_b)$.

\begin{subequations}\label{sys:activesystem}
\begin{align}
 \text{Ownership choices:}\qquad
 \mathcal O&=\{\xi\in[0,1]:G(\xi;I,u_p)\geq0\},\label{eq:systemchoices}\\
 \text{Listed supply:}\qquad
 S&=\int_{\mathcal O}s(\xi)\,dF(\xi)
   =\int_{\mathcal O\cap[0,z(t(I))]}(1-\xi)\,dF(\xi),\label{eq:systemsupply}\\
 \text{Nonowner transportation needs:}\qquad
 V&=\int_{[0,1]\setminus\mathcal O}\xi\,dF(\xi),\label{eq:systemdemand}\\
 \text{Served passenger time:}\qquad
 X&=\beta(I)V,\label{eq:systemflow}\\
 \text{Served trip rate:}\qquad
 D&=\frac Xm,\label{eq:systemtriprate}\\
 \text{Fleet conservation:}\qquad
 S&=I+Dw(I)+X=I+[1+t(I)]X,\label{eq:systemfleet}\\
 \text{Paid utilization:}\qquad
 u_p&=\frac XS,\label{eq:systemutilization}\\
 \text{Committed utilization:}\qquad
 u_b&=[1+t(I)]u_p=1-\frac IS.\label{eq:systemcommitted}
\end{align}
\end{subequations}
Here $\beta(I)$ and $w(I)$ follow \eqref{eq:beta}--\eqref{eq:pickup}, with $t(I)=w(I)/m$. We keep $I$ in the expressions to emphasize their dependence on $I$. The ownership condition \eqref{eq:systemchoices}, together with the optimal listing rule, implements condition 1 of
\Cref{def:equilibrium}. Given these individual choices, \eqref{eq:systemsupply}--\eqref{eq:systemdemand} aggregate owners' listed vehicle time and nonowners' transportation needs, while
\eqref{eq:systemflow}--\eqref{eq:systemtriprate} determine the passenger time served and the corresponding trip rate. These equations implement condition 2 on aggregation. The remaining equations allocate listed vehicle time among idle vehicles, pickup commitments, and passenger service, and determine paid and committed utilization, thereby implementing condition 3 on stationary fleet and utilization consistency. The system thus retains all individual choices and aggregate quantities
in \Cref{def:equilibrium} and expresses its three equilibrium requirements explicitly.

System~\eqref{sys:activesystem} does not admit a closed-form solution, and active equilibria need not be unique. We therefore compute them numerically, using a reduction that makes the search one-dimensional. This tailored search process can, in principle, identify all active equilibria if one exists. The detailed process is provided in Appendix \ref{app:numerical}, and here we only present core idea at a high level. First, to reduce search dimension and improve search coverage, we separate passenger-flow balance from the requirement that the resulting fleet reproduce the assumed idle stock. Combining \eqref{eq:systemflow} and \eqref{eq:systemutilization}, passenger flow balance requires~$u_pS=X=\beta(I)V.$
For a fixed idle stock $I$, we search over paid utilization $u_p$. Let $e$ denote trial paid utilization, and let $S_e(I)$ and $V_e(I)$ denote the supply and nonowner transportation need induced by optimal individual choices based on $I$ and $e$. In Appendix \ref{app:numerical}, we prove that $eS_e(I)=\beta(I)V_e(I)$ has exactly one positive solution $e$ for the given $I$. Denote this solution as $e(I)$. Therefore, passenger flow balance determines a single candidate utilization  $e(I)$ at each idle stock, reducing the remaining equilibrium calculation to a one-dimensional search over $I$. When searching over $I$, we locate zeros of the fleet residual
\[\mathcal Z(I)=S_{e(I)}(I)-I-\beta(I)[1+t(I)]V_{e(I)}(I)\]
which measures the difference between listed supply and the vehicle capacity required for the assumed idle stock, pickup, and passenger on-board time. Every exact zero yields an active equilibrium, and every active equilibrium corresponds to such a zero.

\paragraph{Condition for active equilibrium.} Although the existence of an active equilibrium generally has to be checked numerically, conditions on the primitives are informative in some regimes.

\begin{corollary}[A necessary condition for active equilibrium]\label{co:necessary}
An active equilibrium exists only in a market with $c < 1-v_{\mathrm{ext}}$.
\end{corollary}
\begin{proof}
By definition, an active equilibrium requires positive mass of both owners supplying listed capacity and nonowners requesting service. Its ownership structure must therefore be one of the upper interval or two interval configurations in \Cref{theorem:partition}. In either case, $G(1) = 1-R-c>0$. Since $R>v_{\mathrm{ext}}$, the bound follows as $c<1-v_{\mathrm{ext}}$.
\end{proof}
\cref{co:necessary} indicates that a market with $c \geq 1-v_{\mathrm{ext}}$ does not have an active equilibrium. 
In words, an active market requires that the heaviest users own, and sharing only makes nonownership more attractive to them, so they must already be willing to own without the platform.
A sufficient condition for the existence of an active equilibrium is as follows.
\begin{proposition}[A sufficient condition for active equilibrium]\label{co:sufficient}
Let $t_{\mathrm{thin}}\coloneqq2\bar w_e/(3m)$ denote the limiting pickup ratio as idle stock tends to zero. A market that satisfies
\begin{equation}\label{eq:sufficient}
 a(t_{\mathrm{thin}})>0,\qquad
 \theta_0<z(t_{\mathrm{thin}}),
\end{equation}
is guaranteed to have at least one active equilibrium.
\end{proposition}
This condition requires that some individuals who own in the no-trade equilibrium would also be willing to list their available vehicle time even at the limiting pickup time of a thin market. \cref{app:sufficient} provides the full proof, which uses these incumbent owners to establish a positive core of listed capacity. 
In words, the market can launch from incumbent owners, namely individuals who would own without the platform and who are willing to list even when every match involves the long pickups of a thin market.
The condition is sufficient but not necessary, since investment owners may also supply the initial capacity when listed vehicles are matched almost immediately.
\section{Numerical Studies}\label{sec:Results}

We use the model to study how \gls{acr:p2p} sharing of private \glspl{acr:av} reshapes ownership. 
Our goals are threefold. 
First, we determine when sharing reduces or increases ownership, and who supplies the shared fleet.
Second, we examine how the platform's terms and its protection of owners' access change these outcomes, and whether revenue-maximizing terms reduce ownership.
Third, we assess regulation that targets vehicles bought to be shared.
After the calibration, four case studies move progressively from market structure to platform decisions and to regulation. 


\subsection{Calibration and Computational Protocol}\label{subsec:calibration}
\textit{Parameters.} \Cref{tab:parameters} reports parameters in physical units, with their sources. 
In the model, usage is a fraction of an~$H$-hour operating window, and payoffs are normalized by the value of using one's own vehicle throughout the window.
Durations are therefore divided by~$H$, and monetary values by the value of private vehicle travel, for which we adopt a generalized cost of 1.50\,USD/mile. 
This value combines in-vehicle time valued at about 20\,USD/h, or 0.80\,USD/mile at 25\,mph~\citep{usdot2025bca}, with ownership and operating costs of 0.7\,USD/mile~\citep{aaa2025driving}. 
One unit of usage is then worth $37.5H$\,USD, so daily costs are divided by~$37.5H$ and hourly fares and costs by~37.5. 
For $H=12$ and $\Cav=0$, this gives $c=0.07$, $p=0.5$, $d_o=0.16$, $d_r=0.05$, $v_w=0.8$, $m=0.028$, and $h=0.0069$.
Equilibrium calculations use normalized parameters, and scenarios use the original units for interpretability.

\begin{table}[t]
\centering
\caption{Market parameters of the numerical studies.}
\label{tab:parameters}
\footnotesize
\setlength{\tabcolsep}{3pt}
\begin{tabular}{@{}lllll@{}}
\toprule
Parameter & Baseline & Range explored & Unit & Source\\
\midrule
Operating window $H$ & 12 & $\{24,12,8,4\}$ & h & \\
Mean daily vehicle travel $\bar q$ & 58.6 & & min & \citep{mcguckin2018summary}\\
Type distribution $F_H$ & $\operatorname{Beta}(2,b_H)$ & $b_H\in\{47.1,22.6,14.4,6.19\}^{a}$ & & \citep{kolbl2021physiological}\\
Density $\lambda=M/A$ & 3, 20, 100 & $[1,200]$ & candidates/km$^2$ & \\
Base daily vehicle cost $C_0$ & 31.72 & & USD/day & \citep{aaa2025driving}\\
\gls{acr:av} premium $\Cav$ $^{b}$ & 0, 10, 30 & $[0,90]$ & USD/day & \\
Fare $p$ & 18.75 & $[0.5,35.6]$ & USD/h & \\
Commission $\gamma$ & 0.25 & $[0,0.95]$ & & \citep{schwieterman2020estimating}\\
Owner operating cost $d_o$ & 6 & & USD/h & \citep{aaa2025driving}\\
Renter inconvenience $d_r$ & 1.875 & & USD/h & \\
Waiting cost $v_w$ & 30 & $\{15,30,45,60\}$ & USD/h & \citep{usdot2025bca}\\
External option $v_{\mathrm{ext}}$ & 0 & & & \\
Access-failure penalty $\mu$ & 1.875 & $\{0,0.94,1.875,3.75\}$ & & \\
Trip duration $m$ & 20 & & min & \\
Recall tolerance $h$ & 5 & & min & \\
Dispatch radius $r$ & 2 & & km & \\
Vehicle speed $v$ & 20 & & km/h & \\
\bottomrule
\end{tabular}
\begin{flushleft}\scriptsize
$^{a}$$b_H=2H/\bar q-2$ matches the mean of $H\xi$ to $\bar q$ for $H\in\{24,12,8,4\}$.
$^{b}$ Total \gls{acr:av} cost is $C_0+C_{AV}$.
\end{flushleft}
\end{table}

\begin{figure}[tb]
    \centering
    \includegraphics[width=0.6\linewidth]{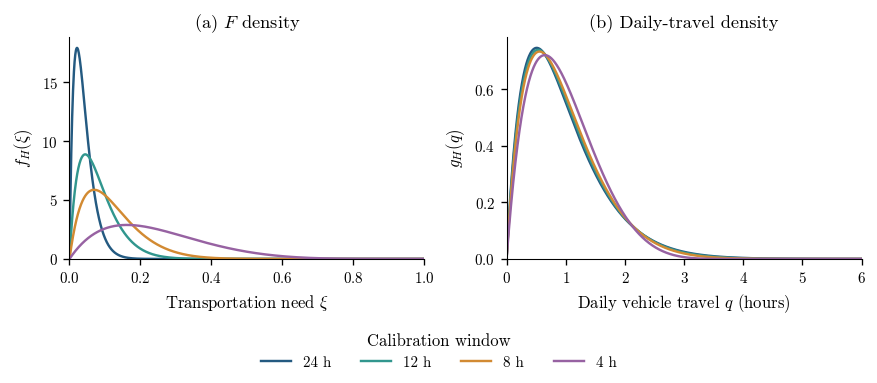}
    \caption{Calibrated type distributions for the four synchronization windows $H$. (a) Density $f_H$ of the usage fraction $\xi\sim\operatorname{Beta}(2,b_H)$. (b) Density of the daily vehicle travel $q=H\xi$, whose mean is fixed.}
    \label{fig:calibration_F}
\end{figure}

\textit{Type distribution and travel synchronization.} 
The 2017 National Household Travel Survey reports an average daily vehicle travel time of 58.6 minutes per person, but no distribution~\citep{mcguckin2018summary}. 
We therefore model the type with a Beta distribution, which is supported on $[0,1]$ as required for a usage fraction. We fix its first shape parameter to 2, which yields the hump-shaped distribution of personal mobility studied by \citet{kolbl2021physiological}.
Since the stationary model does not represent trip timing, we capture the concentration of travel in part of the day through the operating window $H\in\{24,12,8,4\}$, within which all personal travel and all sharing take place. 
A shorter window represents more synchronized travel, which leaves less spare time to share with individuals who travel at other times. 
For each window, $\xi\sim\operatorname{Beta}(2,b_H)$, where $b_H$ matches the mean daily travel $H\xi$ to $\bar q=58.6$ minutes, so the calibrations differ in usage fractions but not in mean travel (\Cref{fig:calibration_F}).


\textit{Representative markets.} 
We label markets by density as \mSparse{} ($\lambda=3$), \mMedium{} ($\lambda=20$), and \mDense{} ($\lambda=100$), and state the premium~$\Cav$ with each market. 
The premium reflects the maturity of \gls{acr:av} technology and, together with the density, determines the ownership structure.

\textit{Computation and reporting.} 
At each parameter point, we compute all active equilibria with the procedure of Appendix~\ref{app:numerical}. 
We scan $I\in[10^{-12}S_{\max},S_{\max}]$ with 400 logarithmic and 200 linear intervals, solve the monotone passenger-flow equation for $u_p$ by bisection, and refine the zeros of the fleet residual~$\mathcal Z(I)$. 
Where several active equilibria coexist, figures show their range, without asserting which one the market selects. 
Unless stated otherwise, ownership changes are relative to the no-trade equilibrium, in percentage points (pp) of the population, and monetary aggregates are in USD per candidate and day. 
Densities are in candidates/km$^2$, premiums and taxes in USD per vehicle and day, and fares and subsidies in USD/h.

\subsection{Case Study I: Does Sharing Reduce the Number of Vehicles?}\label{subsec:cs1}
We compare every active equilibrium with the no-trade benchmark across densities, premiums, and the four synchronization windows, to determine whether sharing reduces ownership and who supplies the fleet.

\begin{figure}[t]
    \centering
    \includegraphics[width=0.8\linewidth]{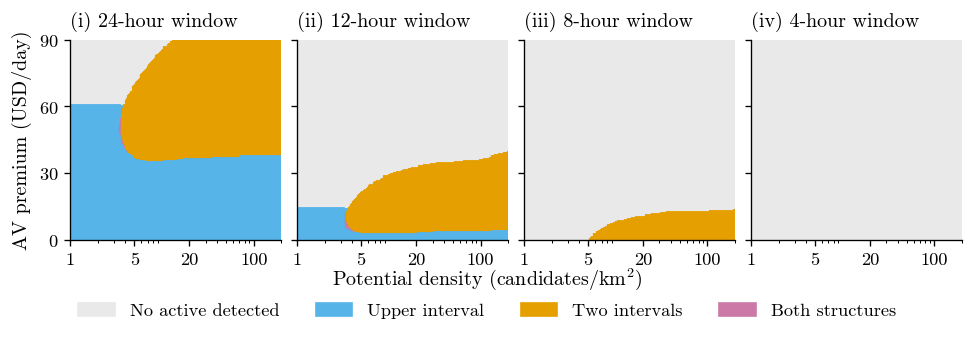}
    \caption{Ownership structure of the detected active equilibria as a function of the potential participant density $\lambda$ and the \gls{acr:av} premium $\Cav$, for the four synchronization windows. Orange: two-interval ownership, with investment owners. Blue: upper-interval ownership. Pink: both structures coexist at different active equilibria. Gray: no active equilibrium is detected.}
    \label{fig:ownership_struct}
\end{figure}

\textit{Where can an active market exist?} 
Travel synchronization determines whether a sharing market can exist at all (\Cref{fig:ownership_struct}). 
No active equilibrium is detected under the 4-hour window, and under the 8-hour window active equilibria exist only in sufficiently dense markets with a low premium, all with two-interval ownership. 
The 24- and 12-hour windows support both structures over a much wider range. A shorter window raises the usage fraction of every type and thus reduces the spare time that owners can list. 
In the calibrated model, a \gls{acr:p2p} market therefore requires that travel be sufficiently staggered over the day.

\textit{Who supplies the shared fleet?} 
Under the 24- and 12-hour windows, two-interval ownership emerges in sufficiently dense markets once the premium exceeds a threshold. 
Higher density improves matching and shortens pickups, so nonowners rely more on the service. 
A higher premium reduces ownership for personal use, which tightens listed supply relative to demand and raises the utilization and income of each listed vehicle. Once this income covers the cost of a vehicle for an individual with almost no personal use, that is, once $u_pa>c$ in \eqref{eq:twointervals}, low-use individuals buy vehicles to share them, and these investment owners can supply a disproportionate share of capacity (details in Appendix~\ref{app:listed}). 
Upper-interval ownership arises instead at low densities, where matching and pickup frictions depress sharing income, or at low premiums, where many individuals own for personal use and list their spare time, so that each listed vehicle earns little. 
In both cases, sharing income does not justify a purchase by low-use individuals, and ownership remains concentrated among high-use individuals.

\begin{figure}[tp]
  \centering
  \begin{subfigure}[b]{0.48\linewidth}
    \centering
    \includegraphics[width=\linewidth]{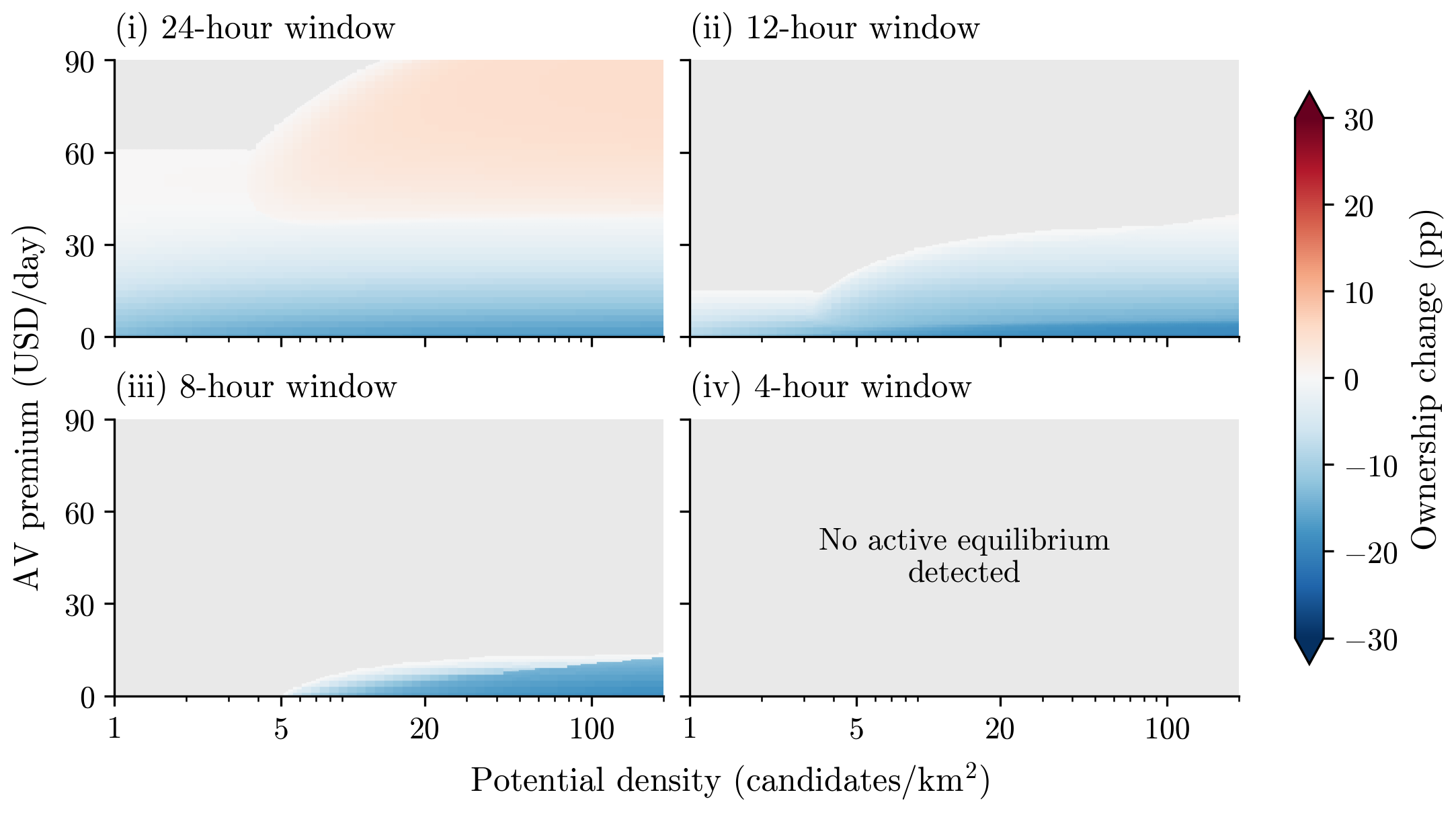}
    \caption{Lowest-ownership active equilibrium.}
    \label{fig:ownership_change_low}
  \end{subfigure}
  \hfill
  \begin{subfigure}[b]{0.48\linewidth}
    \centering
    \includegraphics[width=\linewidth]{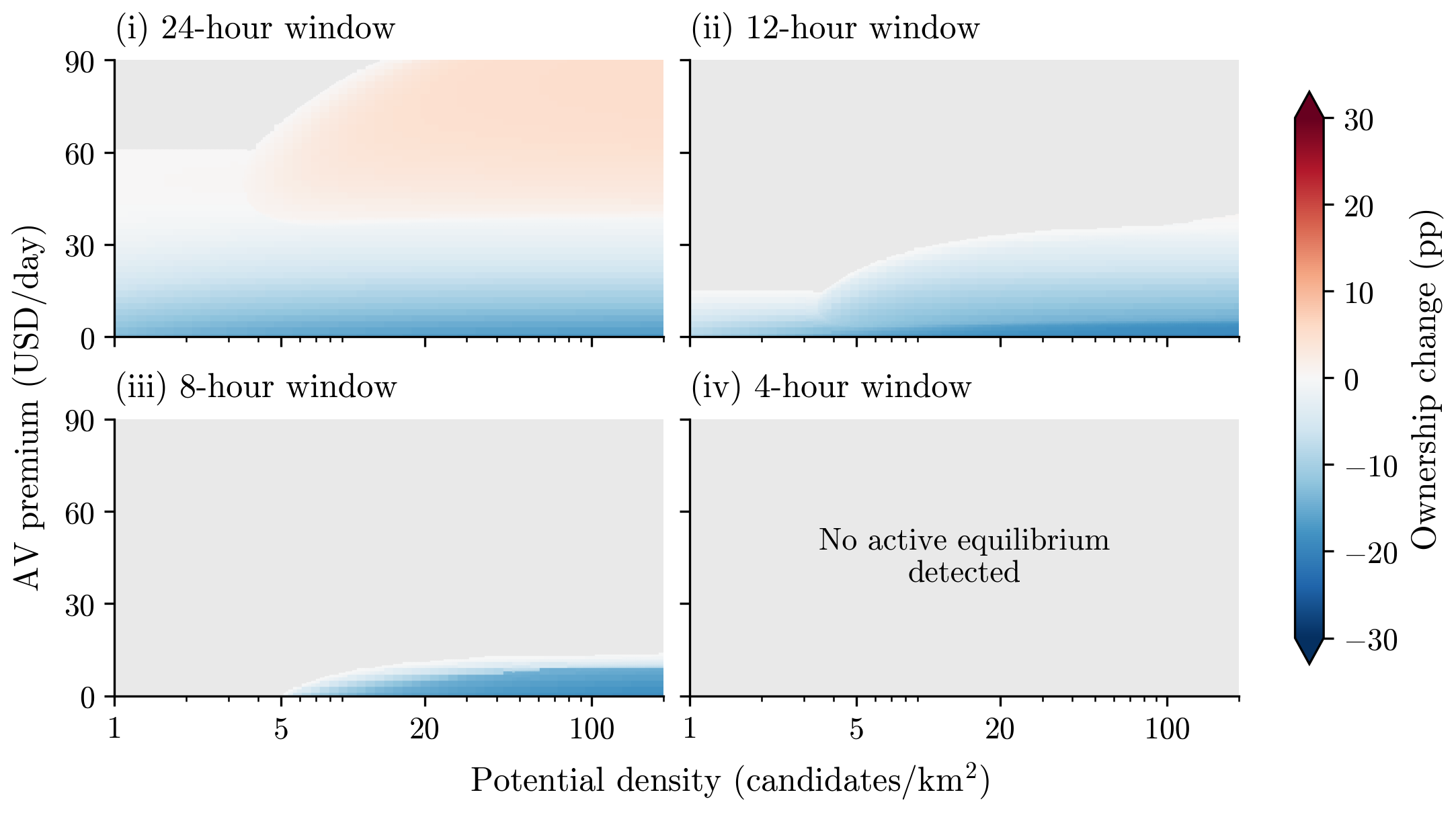}
    \caption{Highest-ownership active equilibrium.}
    \label{fig:ownership_change_high}
  \end{subfigure}
  \caption{Change in the ownership share relative to the no-trade equilibrium, in percentage points, as a function of $\lambda$ and $\Cav$ for the four synchronization windows. Blue: sharing reduces ownership. Red: sharing increases ownership. Gray: no active equilibrium is detected.}
  \label{fig:ownership_change}
\end{figure}

\textit{Does ownership fall?} 
\Cref{fig:ownership_change_low,fig:ownership_change_high} report the lowest and the highest ownership among the detected active equilibria. 
The two maps are almost indistinguishable, so the ownership effect is largely insensitive to equilibrium selection over the tested range. 
The largest gap between coexisting equilibria is 0.85, 0.84, and 11.63\,pp under the 24-, 12-, and 8-hour windows. Appendix~\ref{app:ownership_gap} shows how these gaps vary with market density and the \gls{acr:av} premium in more detail. Under the 24-hour window, sharing reduces ownership when the premium is low and increases it in dense markets with a high premium. 
The increase accompanies two-interval ownership, where the purchases of investment owners outweigh the reduction among individuals who now rely on the service. It also occurs in some upper-interval equilibria, where sharing income makes a purchase attractive to high-use individuals who would not own without the platform, as captured by the balance of listing and access benefits in \eqref{eq:decomposition}.

The 12- and 8-hour windows show the converse, namely that investment ownership need not increase total ownership. Ownership falls in most active markets under the 12-hour window and in all of them under the 8-hour window. In the two-interval regions where it falls, more individuals give up a vehicle than buy one to share it. For example, under the 12-hour window with $\lambda=20$ and $\Cav=10$, ownership falls from 34.40\% to 22.88\%. While 23.96\% of the population give up ownership, investment owners account for only 12.44\%, so ownership falls by 11.52\%.

\textit{What is lost when ownership is held fixed?} 
Existing models of private \gls{acr:av} sharing take the owners as given~\citep{zhang2025park}. 
To quantify the consequences, we fix ownership at $\mathcal O=\mathcal O_0$, let listing and renting adjust, and compute the resulting equilibria, which correspond to the short-run equilibrium of \citet{filippas2020owning}.
On the $61\times46$ density--premium grid, only one of the two models has a detected active equilibrium at 20.9\%, 12.0\%, and 6.3\% of the parameter points under the 24-, 12-, and 8-hour windows, and in every such case it is the model with endogenous ownership. Where both models are active, fixing ownership distorts the scale of service (\Cref{tab:fixed}). 
In \mSparse{}, the benchmark overstates served passenger time and the sharer share and understates utilization. 
In \mMedium{}, it understates served passenger time and utilization and overstates the sharer share. 
In \mDense{}, it has no active equilibrium, whereas investment owners sustain the fleet when acquisition is endogenous. Fixing ownership thus misses both \glspl{acr:av} bought to be shared and the adjustment of spare-time supply and demand among existing participants.

\begin{table}[t]
\centering
\caption{Endogenous (E) versus fixed (F) ownership under the
12-hour window. Entries report E/F. Served passenger time $X$
is measured per candidate as a fraction of the operating window;
sharer shares are percentages of the candidate population.}
\label{tab:fixed}
\small
\setlength{\tabcolsep}{4pt}
\begin{tabular}{@{}lcccc@{}}
\toprule
Market $(\lambda,\Cav)$
& \shortstack{Active market}
& \shortstack{Served demand $X$}
& \shortstack{Sharer share \\ (\%)}
& \shortstack{Utilization $u_p$\\(\%)}\\
\midrule
\mSparse{} $(3,10)$
& Y/Y & 0.013/0.029 & 3.0/9.9 & 47.7/32.2\\
\mMedium{} $(20,0)$
& Y/Y & 0.036/0.020 & 17.1/37.0 & 23.5/6.1\\
\mDense{} $(100,30)$
& Y/N & 0.076/0 & 10.5/0 & 73.4/--\\
\bottomrule
\end{tabular}
\end{table}

Overall, Case Study~I shows that the ownership impact of sharing is decided by the balance between purchases financed by sharing income and the substitution of ownership by the service. Investment owners supply the capacity that makes this substitution possible, so their presence alone does not show that sharing expands the vehicle stock. The net effect depends on how synchronized travel is and, in the few markets with coexisting equilibria, on the equilibrium realized. 
Hereafter, we adopt the 12-hour window as a representative daytime window, which lies between the extremes and supports both structures.

\subsection{Case Study II: How Do Platform Levers Move Ownership}\label{subsec:cs2}
This case study examines the two levers that the platform controls directly, namely its terms and its ability to protect owners' access. First, the fare~$p$ and the commission~$\gamma$ act through three channels. The fare lowers the nonowner surplus $\delta=1-p-d_r-v_{\mathrm{ext}}-v_wt$ and hence~$R$. Second, both change the net sharing income $a=(1-\gamma)p-d_o(1+t)$ and hence the value of owning for a sharer. Third, both also change the utilization~$u_p$, through the numbers of sharers and renters. This third channel makes the response market-specific. Where nonowner needs are already served, a change in the terms redistributes a fixed demand over a fleet of changing size, and where supply is limited, it changes how much travel the service can absorb.

\begin{figure}[t]
    \centering
    \begin{minipage}[c]{0.49\linewidth}
        \centering
        \includegraphics[width=\linewidth]{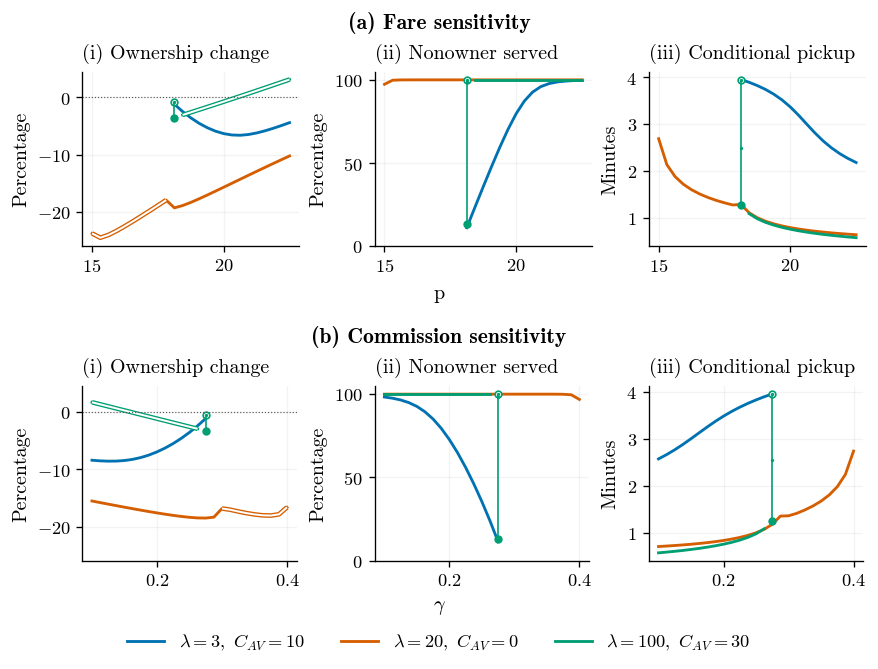}
    \end{minipage}\hfill
    \begin{minipage}[c]{0.49\linewidth}
        \centering
        \includegraphics[width=\linewidth]{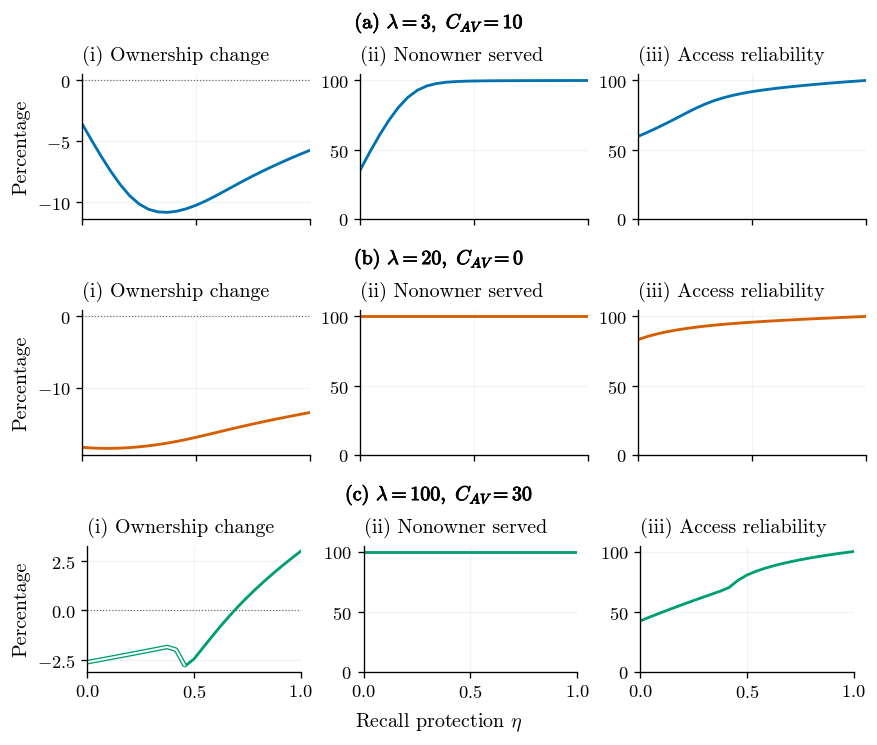}
    \end{minipage}
    \par
    \begin{minipage}[t]{0.49\linewidth}
        \caption{Effect of the fare (a) and the commission (b) in three markets. Columns: ownership change relative to no trade, nonowner demand served, and conditional pickup time. Double and solid lines mark two-interval and upper-interval ownership. Vertical bars span coexisting equilibria, with filled and open circles at their minimum and maximum.}
        \label{fig:sens_f_c}
    \end{minipage}\hfill
    \begin{minipage}[t]{0.49\linewidth}
    \caption{Effect of recall protection $\eta$ in the \mSparse{} (a), \mMedium{} (b), and \mDense{} (c) markets. Columns: ownership change relative to no trade, nonowner demand served, and protected access reliability. In (c)(i), double and solid lines mark two-interval and upper-interval ownership.}
        \label{fig:sens_eta_d}
    \end{minipage}
\end{figure}

\textit{Fare and commission.} \Cref{fig:sens_f_c} reports the ownership change, the served fraction of nonowner needs, and the conditional pickup time as functions of $p$ and $\gamma$ in \mSparse{} with $\Cav=10$, \mMedium{} with $\Cav=0$, and \mDense{} with $\Cav=30$. The full sweeps over $\lambda$ and $\Cav$ are reported in Appendix~\ref{app:sens}.

In \mSparse{}, the response is governed by nonowner access. From the lowest fare with an active equilibrium, a higher fare first attracts supply into the thin market, which raises the served fraction from about 11\% to nearly 100\% and shortens pickups. Living without a vehicle becomes more attractive, so ownership falls despite the higher fare. Once access is nearly complete, the direct effects take over. A higher fare lowers the nonowner surplus and raises sharing income, and ownership increases again. The commission produces the mirror image. A higher commission reduces the served fraction to 10\%, ownership reduction shrinks, and no active equilibrium is detected beyond this point.

In \mMedium{}, access is complete, but the curves show a kink caused by a change of structure. Below $p\approx18$, investment owners supply the entire fleet while high-use owners withhold, and above it high-use owners share their spare time. The transition is governed by the sign of $u_pa-c$ in \eqref{eq:twointervals}, that is, by whether rental income covers the cost of a vehicle for an individual with no personal use. A higher fare raises~$a$ but lowers~$R$, so more nonowners buy a vehicle. Demand shrinks while listed supply expands, utilization falls faster than $a$ rises, and investment ownership ceases to be viable. In a fully served market, a higher fare thus thins the demand on which investment owners depend. The commission produces the opposite transition at $\gamma\approx0.29$. A higher commission lowers~$a$, but marginal high-use owners give up their vehicle, so demand rises and supply contracts until $u_pa$ exceeds~$c$ and investment owners enter.

In \mDense{}, nonowner needs are nearly fully served, except at the lowest fare with active trade, where three equilibria coexist with served fractions of about 13\%, 99\%, and 100\% (vertical bar). A higher fare then offers nonowners nothing and raises the return to owning, so ownership exceeds its no-trade value above $p\approx20.5$. The commission acts in the opposite direction, since a higher commission lowers the net income of owners, and ownership decreases.

Neither a cheaper service nor a lower compensation of owners therefore guarantees fewer vehicles. A lower fare makes the service more attractive but reduces the supply that makes it usable, and can remove the service in a sparse market. A higher commission lowers the income per shared hour, but the resulting withdrawal of supply raises the utilization of the remaining fleet and can make vehicles bought to be shared viable. The direction of the response depends on whether nonowner access is binding, and the composition of ownership can change substantially while the aggregate barely moves.

\textit{Recall protection.} The second lever is operational. A platform that anticipates recalls, for example from owners' schedules, can return the vehicle in time. Let $\eta\in[0,1]$ be the probability that a recall is anticipated and accommodated, so that the mechanism of \eqref{eq:reliability} applies only with probability $1-\eta$. The access reliability and the access-penalty coefficient become ~$\rho_\eta=\eta+(1-\eta)\rho(s)$,~$b_\eta=(1-\eta)\mu(1-h/m+t)$.
Protection thus rescales the access-failure weight from~$\mu$ to $(1-\eta)\mu$, so for $\eta<1$ all results of \Cref{sec:equ} apply and equilibria are computed as in \Cref{subsec:active}. For an owner, protection lowers the access penalty $b_\eta\xi$, which grows with personal use, so it is worth little to low-use owners and most to high-use owners. For a nonowner, it raises~$R$ through the supply that it attracts, until the service saturates. Indirectly, it dilutes the sharing income $u_pa$, because more owners list while fewer individuals rent.

In \mSparse{} (\Cref{fig:sens_eta_d}(a)), the nonowner effect dominates at low protection. As~$\eta$ rises, more needs are served, nonownership becomes more attractive, and the ownership reduction deepens. Once the service saturates, individuals who would otherwise rent now own and share, since sharing is no longer costly for the marginal owner, and the reduction narrows. In \mMedium{} (\Cref{fig:sens_eta_d}(b)), the service is saturated already at $\eta=0$, so protection offers nonowners only slightly shorter pickups. Only the owner-side effect remains, and the ownership reduction narrows. \mDense{} (\Cref{fig:sens_eta_d}(c)) exhibits the structural transition of the fare and commission sweeps. Without protection, investment owners supply the entire fleet while high-use owners withhold. Protection barely affects investment owners since their access penalty is negligible, but high-use types just below~$\theta_H$ begin to own and share, so ownership increases. Their entry shrinks demand and expands capacity, so utilization falls until sharing income no longer covers the cost of a vehicle for a low-use individual, and investment owners disappear at $\eta\approx0.46$. Beyond this point, ownership rises with protection, since sharing income is available at a low access cost, and it exceeds its no-trade level for $\eta\gtrsim0.69$.

Overall, Case Study~II shows that levers that improve the service for one side of the market can increase the number of vehicles. Once nonowner access is not binding, a higher fare, a lower commission, and stronger recall protection all make owning more attractive. In \mDense{}, recall protection turns an ownership reduction of about 2.6\% at $\eta=0$ into an increase of about 3\% at $\eta=1$. Whether a platform lever reduces ownership thus depends primarily on whether nonowner access is binding.

\subsection{Case Study III: Does Platform Revenue Align with Fewer Vehicles?}
\label{sec:cs3}
\begin{figure}[t]
    \centering
    \includegraphics[width=0.8\linewidth]{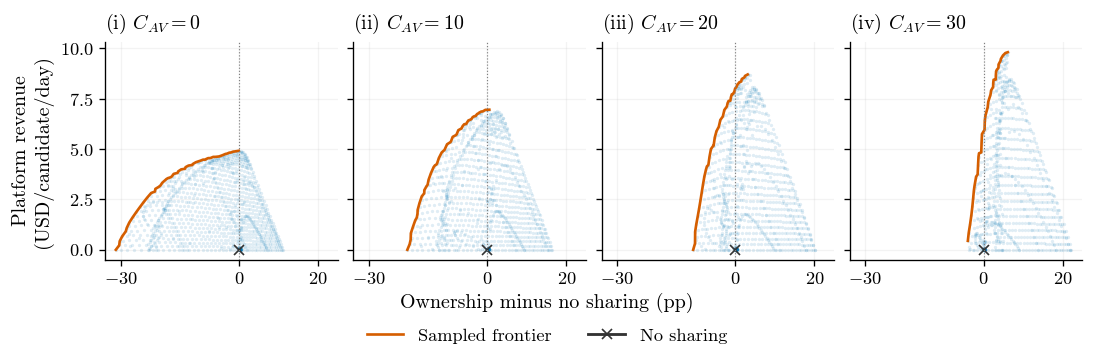}
    \caption{Platform revenue against the ownership change relative to no trade, in the \mMedium{} market ($\lambda=20$) for four \gls{acr:av} premiums. Each dot is a detected active equilibrium for one pair of fare and commission. The orange line is the sampled Pareto frontier, and the cross marks the no-trade equilibrium. }
    \label{fig:tradeoff}
\end{figure}

The fare and the commission that move ownership in Case Studies~I and~II also determine the platform's revenue. This case study asks whether a revenue-maximizing platform reduces ownership. We fix the density at $\lambda=20$ and consider four premiums $\Cav\in\{0,10,20,30\}$ with varying fare over $p\in[0.5,35]$ and the commission over $\gamma\in[0,0.95]$. We compute all detected active equilibria and record the platform revenue $\gamma pX$. \Cref{fig:tradeoff} places each equilibrium in the plane of ownership change and revenue, together with the sampled Pareto frontier of the two objectives.

\textit{Where does a revenue-maximizing platform operate?} Two features are common to all premiums. First, every frontier increases from left to right, so any further ownership reduction along it costs revenue. Second, the revenue maximum lies close to no-trade ownership when vehicles are cheap and above it when they are expensive, by about 6.1\% at $\Cav=30$. The shape of the frontier determines the cost of a reduction. When \glspl{acr:av} are cheap, the frontier is flat near its maximum. At $\Cav=0$, ownership can be reduced by about 10.4\% at the cost of about 10.8\% of the maximal revenue, whereas the full reduction of about 31.2\% eliminates the revenue. As the premium grows, the frontier steepens, and at $\Cav=30$ keeping ownership at its no-trade level already costs about 41.4\% of the maximal revenue.

\begin{figure}[t]
    \centering
    \includegraphics[width=0.6\linewidth]{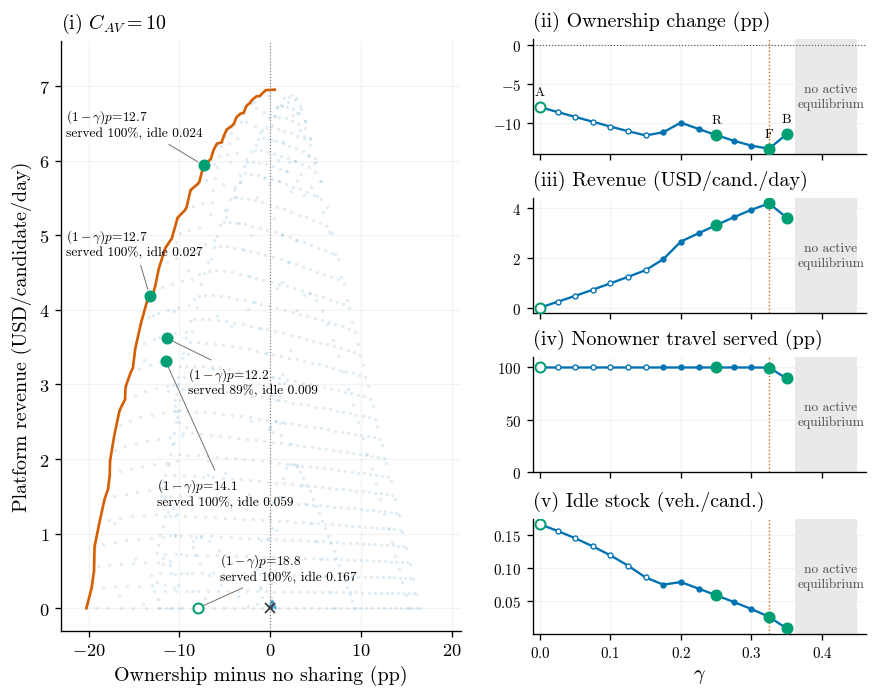}
    \caption{Mechanism behind the frontier at $\Cav=10$. (i) repeats the frontier of \Cref{fig:tradeoff} with five labeled outcomes. Points A, R, F, and B share the fare and differ in the commission, and F$'$ is the frontier outcome with the same owner payout as F at a higher fare. (ii)--(v) trace different metrics at these points. Filled markers denote two-interval ownership and open markers upper-interval ownership.}
    \label{fig:tradeoff_ana}
\end{figure}

\textit{What determines the frontier?} The frontier has a simple interpretation in terms of what owners receive. Along each frontier, fare and commission rise together, while the payout to owners per occupied hour, $(1-\gamma)p$, stays approximately constant. Given this payout, the fare splits the remaining value of served travel between travelers and the platform, and leaving more of it to travelers reduces ownership. \Cref{fig:tradeoff_ana} details the mechanism for $\Cav=10$, and the other premiums behave alike. Points A, R, F, and B share the baseline fare and increase in the commission. At A, owners receive the entire fare, the service is complete, and the platform earns nothing. Ownership falls by only about 8\%, because the ample idle stock keeps utilization, and hence the take-home earnings of owners are low. Raising the commission to R ($\gamma=0.25$) draws down the idle stock and raises revenue, and ownership falls further, because marginal owners give up vehicles that no longer earn enough. At F ($\gamma\approx0.32$), on the frontier, revenue peaks for this fare, the service is still complete, and the idle stock is nearly exhausted (\Cref{fig:tradeoff_ana}(ii)--(v)). At B, listed supply contracts, 11\% of nonowner needs go unserved, revenue falls, and travelers return to ownership, and beyond B no active equilibrium is detected. Point F$'$ pays owners the same 12.7\,USD/h as F at a higher fare and therefore lies higher on the frontier. The frontier thus runs along the edge of capacity exhaustion. Paying owners more forgoes revenue and weakens the ownership reduction, and paying them less collapses the service.

Overall, Case Study~III shows that the platform's incentives are not aligned with fewer \glspl{acr:av}. The revenue-maximizing terms do not reduce ownership in any of our experiments, and they increase it when \glspl{acr:av} are expensive. On the frontier, the platform pays owners just enough to exhaust the idle stock and uses the fare to split the remaining value with travelers. Since the frontier is flat near its maximum when \glspl{acr:av} are cheap, a lower fare at the same payout would buy a substantial ownership reduction at a small revenue cost. A municipality valuing fewer \glspl{acr:av} therefore cannot rely on the platform's incentives, motivating Case Study~IV.

\subsection{Case Study IV: Can Regulation Reduce Ownership?}
\label{sec:cs4}
This case study examines three municipal instruments that act on different sides of the market: a cap on listing time restricts supply and targets investment owners in particular, an ownership tax raises the cost of every \gls{acr:av}, and a passenger subsidy lowers the price that renters pay while the platform and owners still receive their shares of the posted fare. Throughout, we report ownership relative to the unregulated active equilibrium of the same market.

\begin{figure}[t]
    \centering
    \includegraphics[width=0.55\linewidth]{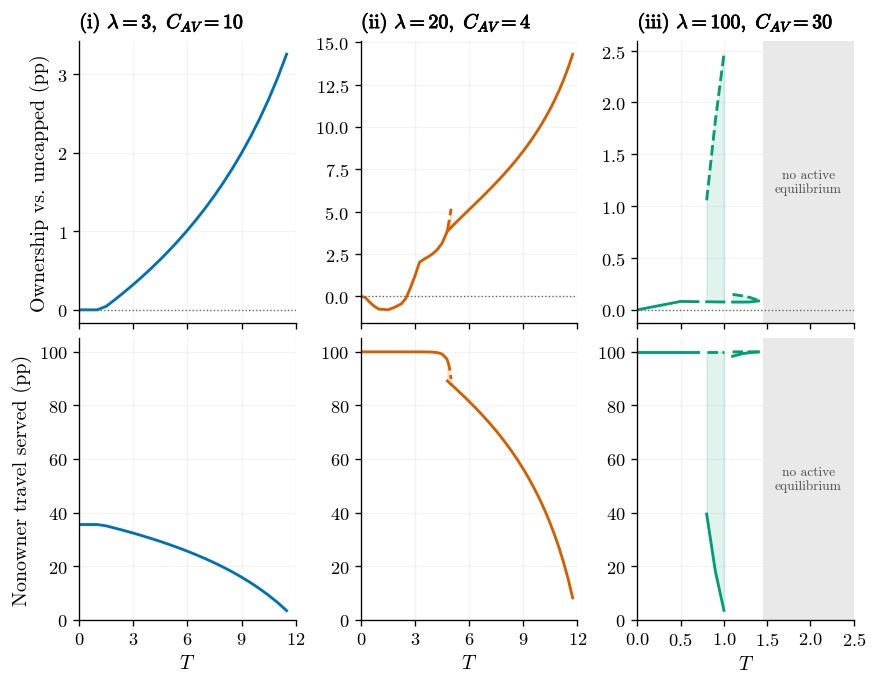}
    \caption{Effect of a daily listing cap in three markets, as a function of the reduction $T$ of the listable time ($T=0$ is uncapped). Top: ownership relative to the uncapped active equilibrium. Bottom: fraction of nonowner travel needs served. Solid and dashed lines bound all detected active equilibria where several coexist, and gray marks caps with no detected active equilibrium.}
    \label{fig:sharing_cap}
\end{figure}

\textit{Listing caps.} A cap limits the listed time of each vehicle to $H-T$ hours per day, so that $s(\xi)\leq\min\{1-\xi,1-T/H\}$, where $T\in[0,H]$ measures how much the cap tightens. \Cref{fig:sharing_cap} reports ownership and the served fraction of nonowner travel as the cap tightens in the three representative markets. In \mSparse{} with $\Cav=10$ (\Cref{fig:sharing_cap}(i)), the fleet consists of high-use owners who share their spare time, so the cap has nothing to target. Once it binds for them, it withdraws capacity from a market that already serves only about a third of nonowner needs, and nonowners return to ownership, which increases by up to about 3.3\%. In \mMedium{} with $\Cav=4$ (column (ii)), investment owners supply part of the fleet and list almost the entire day. A mild cap removes them first and ownership decreases slightly, while high-use sharers absorb the demand without loss of service. Once the cap binds for them too, capacity becomes scarce, the earnings of each listed hour rise, and acquisition becomes worthwhile again, so ownership increases by up to about 14.3\% at the tightest cap. In \mDense{} with $\Cav=30$ (column (iii)), investment owners supply the entire fleet while high-use owners withhold, and no spare-time supply can replace them, so removing about 1.5 hours of listing time eliminates the active equilibrium. Near this threshold, several equilibria coexist, including one with almost no service, so the outcome of a cap is fragile as well as unfavorable.

We observe the same pattern across densities $\lambda\in\{3,20,100,200\}$ and premiums $\Cav\in\{0,2,4,6,10,20,30\}$. A listing cap can lower ownership only where investment owners are present and their supply can be replaced by the spare time of high-use owners. Elsewhere, it only weakens the service that substitutes for ownership. Since the ownership reduction produced by sharing rests on a fleet, a regulation that attacks the fleet undermines the reduction it is meant to protect.

\textit{Ownership tax and passenger subsidy.} A tax~$\Ctax$ adds to the daily cost of every owner and thus acts as an increase of the premium, with $c=(C_0+\Cav+\Ctax)/(37.5H)$. A subsidy~$\psub$ lowers the renters' price to $p-\psub$, while the platform and the owners still receive their shares of~$p$. A subsidy therefore differs from the fare reductions of Case Study~II, which also lower the payout to owners. Per candidate and day, the municipality collects $\Ctax O$, spends $\psub HX$, and runs the net public balance $\Ctax O-\psub HX$.

\begin{figure}[t]
    \centering
    \begin{minipage}[c]{0.49\linewidth}
        \centering
        \includegraphics[width=\linewidth]{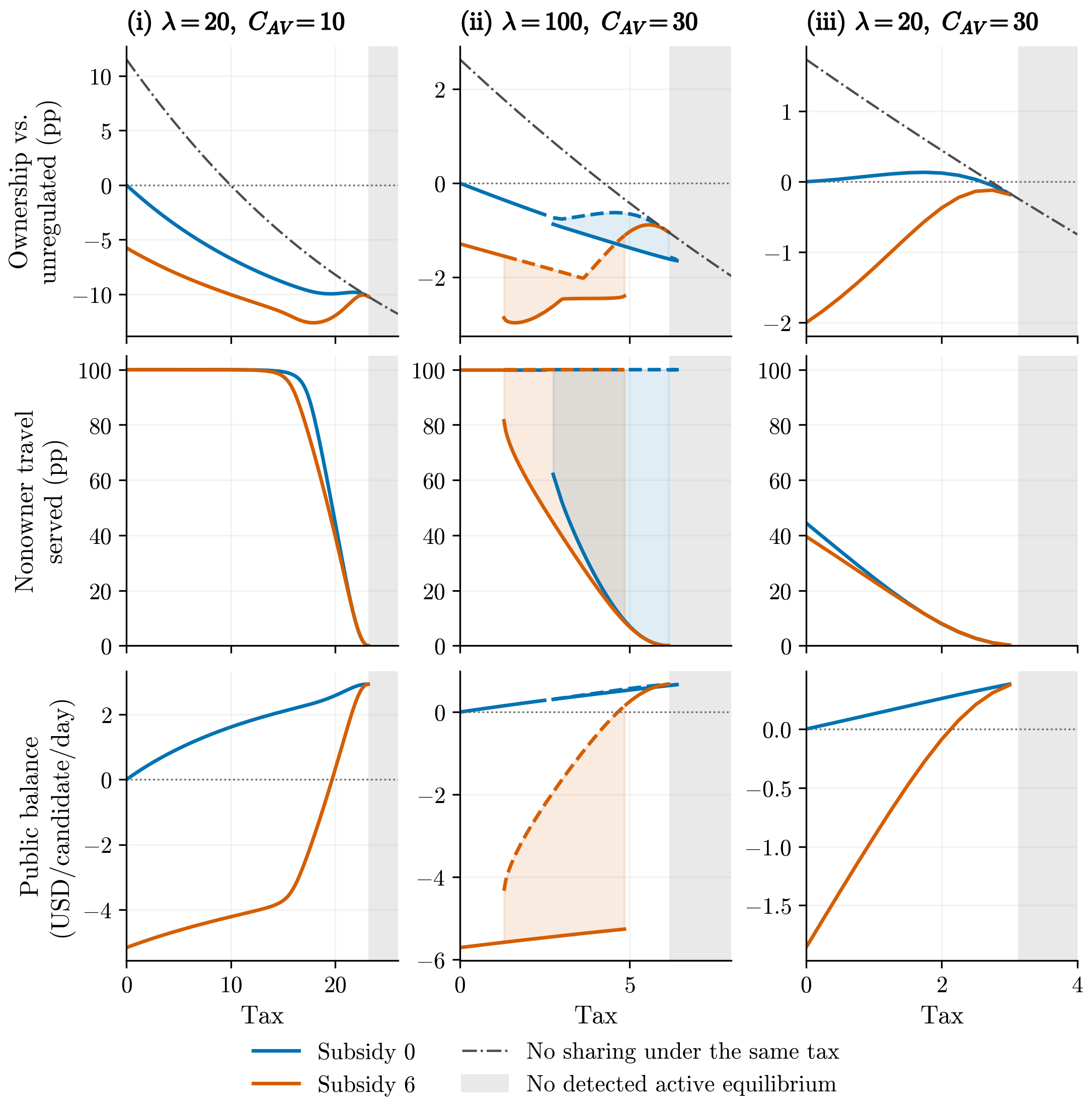}
        \caption{Effect of an ownership tax at two subsidy levels in three markets. Top: ownership relative to the unregulated active equilibrium. The dash-dotted line is ownership without sharing under the same tax, relative to the same reference. Middle: fraction of nonowner travel needs served. Bottom: net public balance.}
        \label{fig:tax}
    \end{minipage}\hfill
    \begin{minipage}[c]{0.5\linewidth}
        \centering
        \includegraphics[width=\linewidth]{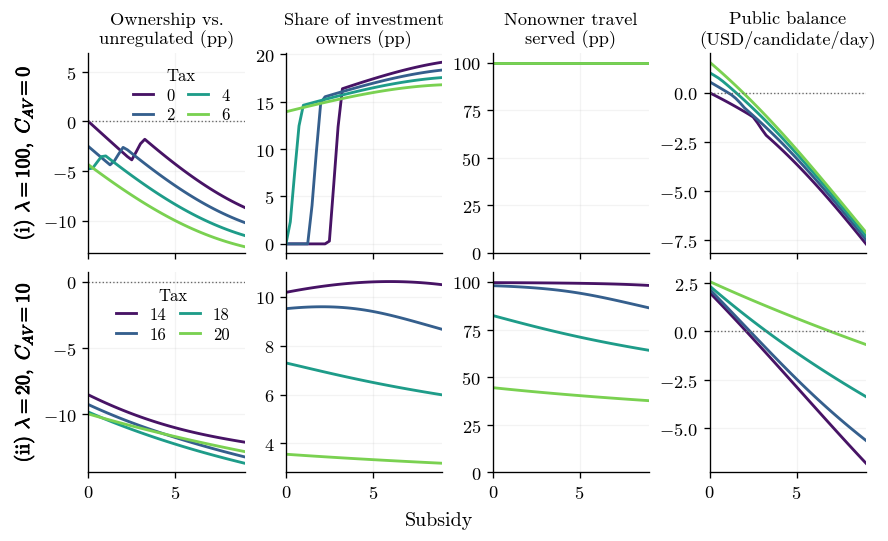}
        \caption{Effect of a passenger subsidy in two markets. Columns: ownership relative to the unregulated active equilibrium, share of investment owners, fraction of nonowner travel needs served, and net public balance. Row (i): \mDense{} market with $\Cav=0$ at $\Ctax$ of 0--6. Row (ii): \mMedium{} market with $\Cav=10$ at $\Ctax$ of 14--20.}
        \label{fig:subsidy}
    \end{minipage}
\end{figure}

A tax lowers ownership even without sharing, since it raises the no-trade threshold \eqref{eq:notrade}. \Cref{fig:tax} therefore compares each market with the same market without sharing under the same tax, so that the gap between the dash-dotted and the solid curves is the ownership effect of sharing. In \mMedium{} with $\Cav=10$ (column (i)), a moderate tax leads marginal owners to give up their vehicles, and the service absorbs nearly all of their travel, so the reduction widens while coverage stays complete. The gap narrows, however, from 11.5\% without tax to about 4.4\% at a tax of 15, so an increasing share of the reduction is due to the tax itself rather than to shared travel. At higher taxes, the service thins rapidly, coverage collapses, and the two curves merge. In \mDense{} with $\Cav=30$ (column (ii)) where investment owners supply the fleet, the tax removes their vehicles. The reduction stalls at a few percent, several equilibria with different service levels coexist, and the active equilibrium disappears at a tax of about 6. In \mMedium{} with $\Cav=30$ (column (iii)), where the service is thin to begin with, a tax of about 3 already removes the active market. Just before, ownership rises slightly above its unregulated level, because the service weakens faster than the tax deters acquisition. This threshold is the same with and without subsidy, so the market disappears on the supply side, where owners can no longer afford their vehicles and the passenger price does not enter.

\Cref{fig:subsidy} shows the effect of the subsidy, which acts through two channels. It makes shared travel cheaper, so nonowners substitute toward the service and ownership falls. It also raises utilization, so sharing earns more and acquisition becomes more attractive. The two rows of \Cref{fig:subsidy} show markets in which different channels prevail. In \mDense{} with $\Cav=0$ (row (i)), the second channel produces a reversal. Ownership first falls, until sharing income covers the cost of a vehicle for an individual with low personal use, that is, until $u_pa>c$. Investment owners then enter, at about 16\% of the population, and ownership jumps up. A larger subsidy lowers ownership again, because high-use owners exit faster than investment owners enter, so part of the subsidy is capitalized into investment fleets. In \mMedium{} with $\Cav=10$ under taxes of 14--20 (row (ii)), the first channel prevails but turns against the service. The subsidy lowers ownership monotonically, but the owners it displaces supply the fleet, so the served fraction falls as the subsidy rises, and more individuals rely on fewer vehicles. At the highest taxes, the subsidies that keep the public balance positive are precisely those under which service erodes.

Overall, Case Study~IV shows that the ownership reduction delivered by sharing lasts only as long as a fleet serves the individuals who gave up their vehicles. Instruments that restrict or burden supply erode this fleet. A listing cap can even increase ownership, and a high tax turns the reduction due to sharing into the conventional effect of a vehicle tax. Instruments that stimulate demand face the opposite limit. In dense markets with cheap vehicles, a subsidy expands service but is partly capitalized by investment owners, whose entry interrupts the ownership reduction. Where vehicles are expensive, the additional demand falls on a shrinking fleet, and service deteriorates. In between, a moderate tax applied while coverage is complete removes marginal owners whose travel the fleet absorbs, so ownership ends up below its level under the same tax without sharing, although the two effects do not add up. In short, policies that make it harder to own the shared fleet make it harder to live without a vehicle.

\section{Conclusion}\label{sec:Conclusion}
In this paper, we studied a \gls{acr:p2p} market in which private \glspl{acr:av} can be listed on an on-demand platform, and in which the option to share feeds back into the decision to own. We developed a stationary equilibrium model that couples ownership, listing, and renting decisions with an aggregate model of spatial matching, pickup, fleet accounting, and recall reliability. We then provided analytical characterizations of the ownership structure in such a sharing market, the definition and categorization of market equilibrium, and a computational procedure for evaluating the resulting market equilibrium outcomes. The analysis shows the importance of considering ownership decisions and shared services together, since the vehicles that make nonownership attractive must themselves be acquired and supplied by someone. By analyzing the computed equilibrium outcomes, our numerical study illustrates how the balance between purchases encouraged by sharing income and ownership displaced by shared vehicles varies with market conditions, and how platform revenue incentives and municipal interventions further shape this balance through their effects on both acquisition and service provision. Concretely, the platform’s revenue-maximizing terms did not reduce ownership in any of our experiments, and regulation aimed at the shared fleet could backfire. The developed model provides a framework to evaluate the impact of these terms and regulations.

Several directions remain open for future research. First, our regulatory study takes the platform's terms as given and examines the corresponding outcomes. A municipality that anticipates the platform's response when deciding on regulatory policies may be more realistically modeled as a leader-follower game. This is a possible future paradigm where the operator of the service and the regulator are two distinct parties. Characterizing this game is thus a natural next step to explore interaction between the two. Second, the stationary model represents the synchronization of travel through the length of an operating window. A time-varying model would capture that owners' needs and the platform's demand peak at the same hours, which lowers both recall reliability and sharing income when they matter most. Third, vehicle rebalancing is not considered in the operation layer of the model. While we explained that \gls{acr:amod} with private \gls{acr:av} is invoked passively only when the vehicle is needed and otherwise parked, the vehicles can still be subject to rebalancing after being invoked, and incorporating rebalancing in operation and fleet accounting enriches the operation that the model can capture. Finally, a reduction in vehicle ownership need not reduce vehicle travel, since shared trips add unoccupied pickups and can attract travelers from transit. Accounting for vehicle hours, congestion, and parking would turn the private-payoff comparison into a welfare analysis that can inform regulation.                                                       


\bibliographystyle{unsrtnat}
\bibliography{references}

\crefalias{section}{appendix}
\crefalias{subsection}{appendix}
\crefalias{subsubsection}{appendix}
\crefname{appendix}{Appendix}{Appendices}
\Crefname{appendix}{Appendix}{Appendices}

\appendix
\section{Spatial matching}\label{app:matching}
Let $L$ be the distance from a passenger to the nearest idle vehicle. The event $L>\ell$ means that the disk of radius $\ell$ contains no idle vehicle. Under the Poisson approximation, the number in this disk has mean $\pi\ell^2(M/A)I$, so
\[
 \Pr(L>\ell)=\exp\{-\pi\ell^2(M/A)I\}.
\]
Successful match requires both a vehicle within the matching radius and an acceptable
pickup time to the passenger. By \eqref{eq:accept}, these requirements are exactly $L\leq r_e=\min\{r,r_A\}$. Rejection occurs before dispatch and uses no vehicle time. Hence the probability of service is
\[
 \Pr(L\leq r_e)=1-e^{-n_eI}=\beta(I),
\]
which proves \eqref{eq:beta}.

\paragraph{Conditional pickup time.}
Write $x=n_eI>0$. Given accepted service, pickup time $P=L/v$ has
conditional distribution
\[
 \Pr(P\leq y\mid L\leq r_e)
 =\frac{1-e^{-x(y/\bar w_e)^2}}{1-e^{-x}},
 \qquad 0\leq y\leq\bar w_e.
\]
The mean of a nonnegative variable is the integral of its survival
probability. Therefore,
\[
 w(I)=\int_0^{\bar w_e}
 \frac{e^{-x(y/\bar w_e)^2}-e^{-x}}{1-e^{-x}}\,dy
 =\bar w_e\frac{\int_0^1e^{-xz^2}\,dz-e^{-x}}{1-e^{-x}},
\]
proving \eqref{eq:pickup}.

\paragraph{Monotonicity and limits.} The following lemma discusses the properties of $\beta$ and $w$.
\begin{lemma}[Monotone matching and pickup]\label{lem:pickup}
The pickup time can be written as~$w(I)=\bar w_e\,\varphi(n_eI)$, where
\begin{equation*}
\varphi(x)\coloneqq\frac{\int_0^1 e^{-xz^2}\,dz-e^{-x}}{1-e^{-x}}.
\end{equation*}
$\varphi$ is continuous and strictly decreasing on $(0,\infty)$, with $\varphi(0^+)=2/3$ and $\lim_{x\to\infty}\varphi(x)=0$. Consequently, $\beta(I)$ is strictly increasing, $w(I)=\bar w_e\varphi(n_eI)$ is strictly decreasing, and $0<t(I)<t_{\mathrm{thin}}=2\bar w_e/(3m)$ for every $I>0$.
\end{lemma}
\begin{proof}
Given $L\leq r_e$, the normalized distance $Y=L/r_e$ has density $f_x(y)=2xye^{-xy^2}/(1-e^{-x})$ on $[0,1]$, and the computation above gives $\varphi(x)=\mathbb E_x[Y]$. For $x'>x$, the likelihood ratio $f_{x'}(y)/f_x(y)$ is proportional to $e^{-(x'-x)y^2}$, which is strictly decreasing in $y$. Hence $Y$ under $x'$ is strictly smaller in the likelihood-ratio order, and therefore in the usual stochastic order, so that $\mathbb E_{x'}[Y]<\mathbb E_x[Y]$. As $x\to0^+$, $f_x(y)\to2y$, the density of the distance from the center of the unit disk to a uniformly drawn point, whose mean is $2/3$. As $x\to\infty$, for every $\varepsilon\in(0,1)$ we have $\Pr_x(Y>\varepsilon)\leq e^{-x\varepsilon^2}/(1-e^{-x})\to0$, and since $Y\leq1$, $\limsup_{x\to\infty}\mathbb E_x[Y]\leq\varepsilon$. Finally, $\beta(I)=1-e^{-n_eI}$ is strictly increasing, and the statements on $w$ and $t=w/m$ follow from $w(I)=\bar w_e\varphi(n_eI)$.
\end{proof}

\begin{remark}[Thin markets, thick markets, and sufficient statistics]
\label{rem:thinthick}
\Cref{lem:pickup} shows that $\varphi$ decreases strictly from~$\varphi(0^+)=2/3$ to~$\lim_{x\to\infty}\varphi(x)=0$. In a thin market ($n_eI\to0$), matches become rare and the expected pickup time approaches $2\bar w_e/3$, the mean distance to a point drawn uniformly in the disk of radius $r_e$ divided by $v$. In a thick market ($n_eI\to\infty$), $\beta\to1$ and $w\approx1/\bigl(2v\sqrt{(M/A)I}\bigr)$, which is the expected distance to the nearest point of a Poisson field of intensity $(M/A)I$ divided by $v$, and which no longer depends on $r$. Consequently, the matching technology $(r,v,M/A)$ affects the equilibrium only through two dimensionless quantities, the market thickness $n_e$ and the ratio $\tau\coloneqq\bar w_e/m$ between the maximum pickup time and the mean trip duration.
\end{remark}

\section{Proof of the population partition}\label{app:partition}
\begin{proof}
We first establish the shape of the ownership gain and then identify its
possible crossings of zero. Since $u_p>0$ and $b>0$, the sharing branch
in \eqref{eq:ownergain} is strictly convex, while the withholding branch
is affine with positive slope $1-R$. Their maximum $G$ is thus continuous
and convex. Therefore, the nonowner set $\{\xi:G(\xi)<0\}$ is an interval,
possibly empty, and owners occupy at most two intervals attached to the
endpoints of $[0,1]$.

If $a\leq0$, no type strictly benefits from listing and
$G(\xi)=(1-R)\xi-c$. This increasing function gives either an upper
ownership interval or no positive mass of owners, as stated in cases 2
and 4. We can therefore restrict attention to $a>0$, for which
\[
 G(0)=u_pa-c,\qquad G(1)=1-R-c.
\]
Feasibility of the operation state gives $u_p(1+t)<1$, and hence
\[
 u_pa<\frac{a}{1+t}
 =\frac{(1-\gamma)p}{1+t}-d_o<(1-\gamma)p\leq p.
\]
Together with \cref{assump:admi}, this orders the endpoints:
\begin{equation}\label{eq:endpoint}
 G(1)-G(0)=1-R-u_pa\geq p+d_r-u_pa>0.
\end{equation}

If $\min G\geq0$, every type weakly prefers ownership, giving case 1.
If $G(1)\leq0$, convexity and the endpoint ordering imply, for $\xi<1$,
\[
 G(\xi)\leq(1-\xi)G(0)+\xi G(1)<0.
\]
There is therefore no positive mass of owners, giving case 4.

In the remaining cases, $G(1)>0$ and $\min G<0$.
If $G(0)\leq0$, the negative interval extends to the lower endpoint,
except for a possible tie at zero, and ends at one interior crossing
$\theta$. Owners form $[\theta,1]$, giving case 2.
If $G(0)>0$, both endpoints are positive and the negative interval lies
strictly between them. Its boundaries satisfy
$0<\theta_L<\theta_H<1$, and owners form
$[0,\theta_L]\cup[\theta_H,1]$, giving case 3.
These cases are mutually exclusive and exhaustive. Neither the strictly
convex quadratic sharing branch nor the increasing affine withholding
branch can vanish on an interval. There are therefore only finitely many
indifferent types, which have zero mass because $F$ is continuous.

The listing rule follows from \Cref{prop:listing}. In case 3,
$\theta_L$ is a downward crossing of $G$, where ownership gives way to
nonownership as personal use increases. This cannot occur on the
increasing withholding branch or at its left endpoint $z$. Thus
$\theta_L<z$, and all lower-interval owners list their available time.
\end{proof}

\section{Equilibrium analysis and computation}\label{app:equilibrium}

\subsection{Relationship between listed supply and trade} \label{app:supply}
\begin{proof}[Proof of \cref{co:listed}]
    To see the first fact, if $D>0$, $Dm=\beta V>0$. Since $\beta$ depends on $I$, \eqref{eq:beta} implies $I>0$, and fleet accounting \eqref{eq:conservation} gives $S=I+D(w+m) \geq Dm>0$. To see the second fact, assume that $S>0$ when $D=0$, then $u_p=u_b=0$ by definition. Fleet accounting in \eqref{eq:committed} gives $I=S>0$. In this case, the optimal owner payoff is $\xi-c$. Comparing the payoffs between nonowner and owner gives
    \[ \xi R-(\xi-c)=c-\xi(1-R)>0 \quad \Rightarrow \quad \xi<\frac{c}{1-R}.\]
    Since $c>0$ and $R<1$, the interval $(0,\min\{1,c/(1-R)\})$ has positive length and consists entirely of types who strictly prefer nonownership. The aggregate transportation need of these types therefore satisfies $V>0$. But $I>0$ gives $\beta>0$, so $Dm=\beta(I)V>0$, which results in a contradiction. Thus $D=0$ implies $S=0$.
\end{proof}

\subsection{Ownership structure conditions}\label{app:ownership}
\begin{proof}[Proof of \Cref{prop:upper_sufficient}]
At an active equilibrium, positive supply requires $a>0$, and fleet
conservation gives $u_p(1+t)<1$. Therefore,
\begin{equation}\label{eq:incomebound}
 u_pa<\frac{a}{1+t}
 =\frac{(1-\gamma)p}{1+t}-d_o
 \leq(1-\gamma)p-d_o.
\end{equation}
If $c\geq(1-\gamma)p-d_o$, it follows that $G(0)=u_pa-c<0$.
Investment ownership is excluded, so the active classification
\eqref{eq:upper}--\eqref{eq:twointervals} implies upper-interval ownership.

For the second condition, we show that ownership becomes more attractive
as personal use increases. By \cref{assump:admi}, $1-R\geq p+d_r$.
Also,
\[
 u_pb=\mu u_p(1-h/m+t)\leq\mu u_p(1+t)<\mu.
\]
On the sharing branch, these bounds and \eqref{eq:incomebound} give
\[
 G'(\xi)=1-R-u_p(a+b)+2u_pb\xi
 >\gamma p+d_o+d_r-\mu.
\]
If $\mu\leq\gamma p+d_o+d_r$, this derivative is positive.
The withholding branch also has positive slope $1-R$, and the two
branches join continuously. Thus $G$ is strictly increasing, which
rules out low-use owners separated from high-use owners by nonowners.
An active equilibrium must again have upper-interval ownership.
\end{proof}

\begin{proof}[Proof of \Cref{co:two_necessary}]
A two-interval active equilibrium cannot satisfy either sufficient
condition in \Cref{prop:upper_sufficient}. Hence
$c<(1-\gamma)p-d_o$ and $\mu>\gamma p+d_o+d_r$.
Moreover, investment entry requires $c<u_pa$. Combining this with the
strict utilization bound in \eqref{eq:incomebound} yields
\[
 c<\frac{(1-\gamma)p}{1+t}-d_o.
\]
Multiplying by $1+t>0$ and dividing by $c+d_o>0$ gives
$t<(1-\gamma)p/(c+d_o)-1$, as claimed.
\end{proof}

\subsection{Individual payoff comparison}\label{app:payoff}
\begin{proof}[Proof of \Cref{prop:payoff}]
At no trade, type $\xi$ chooses between owning for private use and using
the external option, so the optimized payoff is
\[
 U^0(\xi)=\max\{\xi-c,\ \xi v_{\mathrm{ext}}\}.
\]
At an active equilibrium, optimal listing and passenger acceptance give
\[
 U^*(\xi)=\max\left\{
 \xi-c+u_p(1-\xi)[a-b\xi]^+,\quad
 \xi v_{\mathrm{ext}}+\xi\beta\delta\right\},
\]
where $[q]^+=\max\{q,0\}$. Each alternative is at least as valuable as
its no-trade counterpart: an owner can withhold, while a nonowner can
use the outside option whenever sharing is unavailable or unattractive.
Thus $U^*(\xi)\geq U^0(\xi)$ for every type, including those whose
ownership decision changes.

For strict improvement, take
$0<\xi<\min\{\theta_0,1\}$. Such a type strictly prefers nonownership
at no trade, so $U^0(\xi)=\xi v_{\mathrm{ext}}$.
At active trade, $\beta>0$. Since $r_e\leq r_A$, we have
$v_w\bar w_e/m\leq\Delta_0$, while \Cref{lem:pickup} gives
$t<2\bar w_e/(3m)$. Hence $\delta=\Delta_0-v_wt>\Delta_0/3>0$.
Consequently,
\[
 U^*(\xi)\geq\xi v_{\mathrm{ext}}+\xi\beta\delta>U^0(\xi).
\]
Since $c>0$, this interval has positive length and, by full support of
$F$, positive population mass.
\end{proof}

\subsection{Equilibrium search algorithm}\label{app:numerical}
System~\eqref{sys:activesystem} jointly determines individual choices,
aggregate supply and demand, and fleet utilization. Its nonlinear dependence
on idle vehicle availability generally prevents a closed-form solution, so
we solve it numerically. A standard approach is to express the equations as
residuals and use a numerical root solver from multiple trial points.
A grid can distribute these starting points across the search domain, but
evaluating a grid alone does not solve the equations: candidate solutions
must be refined and checked against the original system. Moreover, the
number of grid points grows rapidly with the number of independently
searched variables.

For \eqref{sys:activesystem}, we can substantially reduce this search.
Given idle stock $I$ and paid utilization $u_p$, individual choices determine supply and demand, and several other quantities follow directly from their
definitions. This leaves two balance equations in two unknowns. We then
establish that, for every fixed idle stock, passenger-flow balance determines
exactly one trial utilization. Substituting that utilization into fleet
balance leaves a single equation in idle stock. We next show the system reduction process and conditional uniqueness result.

\paragraph{From the full system to two balance equations.}
Fix a trial idle stock $I>0$ for which $a(t(I))>0$, as positive sharing
returns are necessary for active supply. Equations~\eqref{eq:beta}--\eqref{eq:pickup}
determine matching probability $\beta(I)$ and conditional pickup time $w(I)$.
These determine
\[
\begin{aligned}
 t(I)&=w(I)/m,\\
 a(I)&=(1-\gamma)p-d_o[1+t(I)],\\
 b(I)&=\mu[1-h/m+t(I)],\\
 R(I)&=v_{\mathrm{ext}}+\beta(I)[\Delta_0-v_wt(I)],\\
 z(I)&=\min\{1,a(I)/b(I)\}.
\end{aligned}
\]
Here $a(I)$, $b(I)$, and $z(I)$ abbreviate $a(t(I))$, $b(t(I))$, and
$z(t(I))$, respectively.

The ownership gap also depends on paid utilization. Let $e>0$ be a trial
value of that utilization. Given $(I,e)$, calculate
\begin{equation}\label{eq:trialgap}
 G_e(\xi;I)=\xi[1-R(I)]-c
 +e(1-\xi)[a(I)-b(I)\xi]^+.
\end{equation}
The owner and nonowner sets are then
\[
 \mathcal O_e(I)=\{\xi:G_e(\xi;I)\geq0\},\qquad
 \mathcal N_e(I)=[0,1]\setminus\mathcal O_e(I).
\]
This implements the ownership condition~\eqref{eq:systemchoices}.
The optimal listing rule gives full listing by owners below $z(I)$ and
withholding by owners above it. Thus \eqref{eq:systemsupply}--\eqref{eq:systemdemand}
determine
\begin{equation}\label{eq:trialstocks}
 S_e(I)=\int_{\mathcal O_e(I)\cap[0,z(I)]}(1-\xi)\,dF(\xi),\qquad
 V_e(I)=\int_{\mathcal N_e(I)}\xi\,dF(\xi).
\end{equation}
These quantities are calculated from individual choices; they do not require
independent searches. We next use the remaining equations sequentially.

First, \eqref{eq:systemflow} determines served passenger time:
\[
 X_e(I)=\beta(I)V_e(I).
\]
Thus $X$ is determined by matching probability and nonowner demand.
Next, \eqref{eq:systemtriprate} determines the served trip rate:
\[
 D_e(I)=\frac{X_e(I)}m=\frac{\beta(I)V_e(I)}m.
\]
Consequently, both passenger-on-board time $X_e(I)$ and pickup vehicle time
$D_e(I)w(I)$ are known at the trial pair.

Equation~\eqref{eq:systemfleet} imposes fleet conservation. The supply
generated by individual choices must accommodate the assumed idle stock
and these two service commitments:
\[
 S_e(I)=I+D_e(I)w(I)+X_e(I).
\]
Substituting the quantities determined by \eqref{eq:systemflow} and
\eqref{eq:systemtriprate}, and using $t(I)=w(I)/m$, gives
\[
 S_e(I)=I+[1+t(I)]\beta(I)V_e(I).
\]
This equality need not hold at an arbitrary trial pair $(I,e)$. It is
therefore one of the two conditions that remain to be solved.

Equation~\eqref{eq:systemutilization} requires consistency of paid
utilization. At an active equilibrium, the trial utilization $e$ must equal
the passenger-on-board fraction of listed supply:
\[
 e=\frac{X_e(I)}{S_e(I)}.
\]
Substituting \eqref{eq:systemflow} and multiplying by $S_e(I)>0$ gives
\[
 eS_e(I)=\beta(I)V_e(I).
\]
This is the second remaining condition: passenger service implied by
utilization and listed supply must equal the service generated by matched
nonowner demand.

Finally, \eqref{eq:systemcommitted} determines committed utilization without
adding an independent equation. Its first equality gives
$u_b=[1+t(I)]e$. When the two preceding balance conditions hold,
\[
 [1+t(I)]e
 =\frac{[1+t(I)]\beta(I)V_e(I)}{S_e(I)}
 =\frac{S_e(I)-I}{S_e(I)}
 =1-\frac{I}{S_e(I)}.
\]
Thus its second equality follows automatically from fleet conservation
and paid-utilization consistency.

The full system therefore reduces to two equations in the trial variables
$(I,e)$:
\begin{subequations}\label{eq:twobalances}
\begin{align}
 eS_e(I)-\beta(I)V_e(I)&=0,
 &&\text{passenger-flow balance},\label{eq:trialflow}\\
 S_e(I)-I-[1+t(I)]\beta(I)V_e(I)&=0,
 &&\text{fleet balance}.\label{eq:trialfleet}
\end{align}
\end{subequations}
All other quantities are determined from $(I,e)$ through the preceding
construction. We next show that the first equation determines a single
$e(I)$ at each fixed $I$, leaving only the second equation to be solved
over idle stock.

\paragraph{Conditional uniqueness and the scalar reduction.}
At a fixed idle stock, increasing paid utilization raises the return from
ownership for individuals willing to share. Listed supply therefore cannot
decrease, while nonowner transportation demand cannot increase. These
responses imply that passenger-flow balance has a unique solution.

\begin{lemma}[Conditional utilization and equilibrium recovery]\label{lem:reduction}
For every $I>0$ with $a(t(I))>0$, equation~\eqref{eq:trialflow} has exactly
one solution $e(I)>0$, satisfying
\begin{equation}\label{eq:innerbalance}
 e(I)S_{e(I)}(I)=\beta(I)V_{e(I)}(I).
\end{equation}
The fleet residual
\begin{equation}\label{eq:fleetresidual}
 \mathcal Z(I)=S_{e(I)}(I)-I
 -[1+t(I)]\beta(I)V_{e(I)}(I)
\end{equation}
is continuous on this domain. Its zeros correspond exactly to active
equilibria, with $u_p=e(I)$.
\end{lemma}

\begin{proof}
\emph{Existence and uniqueness at fixed idle stock.}
Fix $I>0$ with $a(t(I))>0$, and define
\[
 f_I(e)=eS_e(I)-\beta(I)V_e(I).
\]
Increasing $e$ expands the ownership set, so $S_e(I)$ is
nondecreasing and $V_e(I)$ is nonincreasing. Both aggregates
are jointly continuous in $(I,e)$: the ownership gap is continuous,
and its finitely many indifferent types have zero mass under $F$.
Thus $f_I$ is continuous and nondecreasing, and is strictly
increasing wherever $S_e(I)>0$.

At $e=0$, $G_0(0;I)=-c<0$, so an interval of low-use
types strictly prefer nonownership. Full support of $F$
implies $V_{e=0}(I)>0$; since $\beta(I)>0$, we obtain
$f_I(0)=-\beta(I)V_{e=0}(I)<0$. For sufficiently large
$e$, every type in a closed interval strictly inside $(0,z(I))$
owns and lists, because the listing return is bounded away
from zero there. Hence supply is bounded below by a positive
constant, while demand remains bounded, so $f_I(e)\to\infty$.

Continuity and the opposite signs give a positive zero.
Any flat portion of $f_I$ occurs before supply becomes
positive, where $f_I<0$. Once supply is positive, $f_I$
is strictly increasing, so the zero is unique.

\emph{Continuity as idle stock varies.}
Joint continuity of $f_I(e)$ and its unique sign-changing
zero imply continuity of $e(I)$: trial values just below
and above the zero retain opposite residual signs for
nearby idle stocks. Continuity of the aggregates and
service quantities then implies continuity of $\mathcal Z(I)$.

\emph{Recovering equilibrium.}
At a zero of $\mathcal Z$, write $S=S_{e(I)}(I)$ and $V=V_{e(I)}(I)$.
Both balance equations hold:
\[
 eS=\beta V>0,\qquad S=I+(1+t)\beta V.
\]
Combining them gives
\[
 0<(1+t)e=1-\frac IS<1.
\]
Thus utilization is physically feasible. The preceding construction
already satisfies individual choices, aggregation, and the definitions
of served passenger time and trip rate. The zero therefore recovers an
active equilibrium of \eqref{sys:activesystem}. Conversely, every active
equilibrium satisfies both balance equations. Conditional uniqueness
implies $u_p=e(I)$, and fleet conservation implies $\mathcal Z(I)=0$.
\end{proof}

The lemma removes utilization as an independent search dimension.
Instead of searching a two-dimensional grid of $(I,e)$, we search over
$I$ and solve a unique scalar passenger-flow equation at each trial stock.
This inner calculation still requires numerical evaluations, but its
monotone crossing permits a bracketed method such as bisection.
The remaining equilibrium problem is
\[
 \mathcal Z(I)=0.
\]
The residual need not be monotone in $I$, so several active equilibria
may remain.

At an active equilibrium, positive idle stock requires committed utilization
to be strictly below one. Hence paid utilization must satisfy
\[
 0<e<\bar e(I):=\frac{1}{1+t(I)}.
\]
This bound provides a finite bracket for the inner search and allows
infeasible trial idle stocks to be discarded before solving for utilization.
Recall that
\[
 f_I(e)=eS_e(I)-\beta(I)V_e(I),\qquad f_I(0)<0.
\]
By \Cref{lem:reduction} and the sign pattern established in its proof,
the unique inner solution lies strictly below $\bar e(I)$ exactly when
\[
 f_I(\bar e(I))>0.
\]
If this inequality holds, bisection on $[0,\bar e(I)]$ finds the unique
feasible $e(I)$. Otherwise, the unique inner solution lies at or above
the physical bound, so this idle stock cannot support an active equilibrium
and can be discarded. Equality is also excluded: $e=\bar e(I)$ implies
committed utilization equal to one, leaving no capacity for the positive
idle stock.

\paragraph{Complete search procedure.}
Every active equilibrium satisfies
\begin{equation}\label{eq:searchdomain}
 0<I<S_{\max}:=\int_0^1(1-\xi)\,dF(\xi),\qquad a(t(I))>0,
\end{equation}
because idle stock is strictly smaller than listed supply, and listed
supply cannot exceed the time available if everyone owned and listed
all capacity remaining after personal use. $S_{\max}$ therefore provides an upper bound of the search region. The numerical procedure is
as follows.
\begin{enumerate}
 \item \emph{Choose trial idle stocks.} Construct a grid over
 \eqref{eq:searchdomain}, using logarithmic spacing near zero and additional
 points over the remaining interval. Record the positive numerical lower cutoff.
 \item \emph{Calculate service conditions.} At each trial $I$, calculate
 $\beta,w,t,R,a,b,z$.
 \item \emph{Solve passenger-flow balance.} At the trial idle stock $I$,
 set $\bar e(I)=1/[1+t(I)]$. For each utilization evaluated, calculate
 the ownership gap $G_e$, identify the owner and nonowner sets, and
 integrate to obtain $S_e$ and $V_e$. The ownership boundaries follow
 from the quadratic sharing branch and the affine withholding branch
 on their respective domains. First evaluate $f_I(\bar e(I))$.
 If it is nonpositive, discard this idle stock and continue to the next
 trial $I$. If it is positive, use bisection on $[0,\bar e(I)]$, with
 $f_I(0)<0$, to obtain the unique feasible utilization $e(I)$.
 Proceed to the fleet-balance calculation only for trial idle stocks
 that pass this check.
 \item \emph{Evaluate fleet balance.} Substitute $e(I)$, $S_{e(I)}(I)$,
 and $V_{e(I)}(I)$ into $\mathcal Z(I)$. Refine each sign-changing interval
 in $I$, resolving the inner problem whenever a new idle stock is evaluated.
 Examine sampled local extrema for closely spaced roots or possible tangencies.
 \item \emph{Recover and verify equilibria.} At each candidate root $I^*$, set
 \[
 \begin{aligned}
 u_p&=e(I^*),& S&=S_{e(I^*)}(I^*),& V&=V_{e(I^*)}(I^*),\\
 X&=\beta(I^*)V,& D&=X/m,& u_b&=[1+t(I^*)]u_p.
 \end{aligned}
 \]
 Recover ownership and listing from the corresponding individual choices.
 Check all equations in \eqref{sys:activesystem}, marginal indifference,
 utilization bounds, and recall reliability. Retain every verified active
 equilibrium and compute no trade separately.
 \item \emph{Check numerical sensitivity.} Repeat the search with finer
 grids and report the search cutoff, numerical tolerances, and equilibrium
 residuals.
\end{enumerate}
The exact scalar characterization includes every active equilibrium.
A finite numerical search does not certify exhaustive detection: closely
spaced roots, tangent roots, and roots below the numerical cutoff may be
missed. Conditional uniqueness therefore reduces the search dimension
without implying uniqueness of the market equilibrium.

\paragraph{Equivalent bounded inner calculation.}
The experimental implementation uses a bounded alternative to step~3.
Eliminating supply between \eqref{eq:trialflow} and \eqref{eq:trialfleet} gives
\begin{equation}\label{eq:boundedinner}
 eI=\beta(I)V_e(I)[1-(1+t(I))e],\qquad
 0\leq e\leq\frac{1}{1+t(I)}.
\end{equation}
The left side strictly increases with $e$, while the right side is
nonnegative and nonincreasing. At zero the left side is smaller; at the
upper endpoint it is larger. Thus there is a unique interior solution
$\widehat e(I)$, which can be found by bisection on this fixed interval.

Evaluate the fleet residual using $\widehat e(I)$, and denote it by
$\widehat{\mathcal Z}(I)$. Equation~\eqref{eq:boundedinner} implies
\[
 \widehat e(I)S_{\widehat e(I)}(I)-\beta(I)V_{\widehat e(I)}(I)
 =\widehat e(I)\widehat{\mathcal Z}(I).
\]
A zero of this residual therefore also satisfies passenger-flow balance.
Conversely, every active equilibrium satisfies \eqref{eq:boundedinner}.
The bounded formulation consequently has exactly the same equilibrium
roots as the preceding procedure, although the two inner solutions and
outer residuals generally differ away from equilibrium. The remaining
search and verification steps apply with $\widehat e$ and
$\widehat{\mathcal Z}$ in place of $e$ and $\mathcal Z$; representative
cases can also be checked using the direct passenger-flow formulation.

\subsection{Sufficient condition for active equilibrium}\label{app:sufficient}
\begin{proof}[Proof of \Cref{co:sufficient}]
Write $z_{\mathrm{thin}}=z(t_{\mathrm{thin}})$.
The proof identifies a group who would own and share even in a thin
market, then uses that group's supply to show that the fleet residual
changes sign.

\emph{A positive core of owner-sharers.}
By \Cref{lem:pickup}, $t(I)<t_{\mathrm{thin}}$. Since $a(t)$
is nonincreasing in $t$, the assumption $a(t_{\mathrm{thin}})>0$
ensures $a(t(I))>0$ for all $I>0$. \Cref{lem:reduction} therefore gives
continuous $e(I)$ and $\mathcal Z(I)$ for every positive idle stock.
Write $S(I)=S_{e(I)}(I)$ and $V(I)=V_{e(I)}(I)$.

Choose a closed interval $J=[\xi_1,\xi_2]$ strictly inside
$(\theta_0,z_{\mathrm{thin}})$, with $\xi_1<\xi_2$.
As $I\downarrow0$, $\beta(I)\to0$, $R(I)\to v_{\mathrm{ext}}$, and
$t(I)\to t_{\mathrm{thin}}$. Thus, for every sufficiently small $I>0$
and every $\xi\in J$,
\[
 \xi[1-R(I)]-c>0,\qquad a(t(I))-b(t(I))\xi>0.
\]
The first inequality makes ownership preferable even without sharing
income; the second makes full listing optimal at any $e(I)>0$.
Consequently,
\[
 S(I)\geq S_J:=\int_J(1-\xi)\,dF(\xi)>0.
\]
Full support of $F$ and $\xi_2<1$ give the strict positivity.

\emph{A change of sign in fleet balance.}
Let $V_{\max}=\int_0^1\xi\,dF(\xi)$. Since $V(I)\leq V_{\max}$,
\[
 \mathcal Z(I)\geq S_J-I-(1+t_{\mathrm{thin}})\beta(I)V_{\max}>0
\]
for sufficiently small $I>0$. Economically, the core owners supply a
positive amount while the vehicle time required for the few accepted
trips tends to zero.

At the other end, supply never exceeds
\[
 S_{\max}:=\int_0^1(1-\xi)\,dF(\xi)=1-\mathbb E[\xi]<1.
\]
This is the listed capacity if everyone owned and listed all time remaining
after personal use. The inequality is strict because full support of $F$
implies positive mean personal use, $\mathbb E[\xi]>0$.
Thus the auxiliary residual at $I=1$ satisfies
\[
 \mathcal Z(1)\leq S_{\max}-1<0.
\]
Continuity and the intermediate value theorem yield
$I^*\in(0,1)$ with $\mathcal Z(I^*)=0$.
By \Cref{lem:reduction}, this zero is an active equilibrium; in particular,
its utilization is feasible and $I^*<S_{\max}$.
The conditions therefore guarantee at least one active equilibrium,
without asserting uniqueness or selecting its ownership structure.
\end{proof}

\section{{Additional Numerical Results}}
\subsection{Listed capacity among owners}\label{app:listed}
\begin{figure}[ht]
    \centering
    \includegraphics[width=0.5\linewidth]{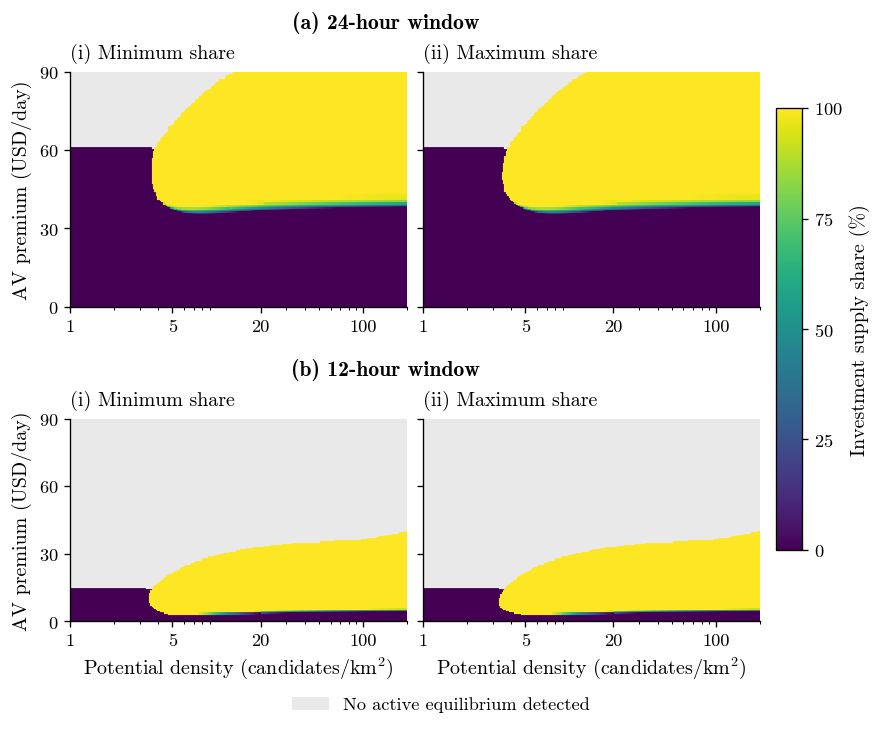}
    \caption{Percentage of total listed vehicle time supplied by investment owners under the 24-hour (a) and 12-hour (b) windows. Within each row, columns (i) and (ii) report the minimum and maximum shares across detected active equilibria at each $(\lambda,\Cav)$.}
    \label{fig:investment_supply}
\end{figure}

\Cref{fig:investment_supply} complements the ownership structure maps in Case Study I by showing how much shared capacity investment owners provide. At a two-interval equilibrium, their share of total listed capacity is
\[
    \frac{\displaystyle\int_0^{\theta_L}(1-\xi)\,dF_H(\xi)}{S},
\]
since all low-use owners list their available vehicle time. This share is zero at an upper interval equilibrium. A substantial part of the supply of the two-interval region is supported by the investment owners.

\subsection{Ownership gap across equilibrium} \label{app:ownership_gap}
\begin{figure}[ht]
    \centering
    \includegraphics[width=0.6\linewidth]{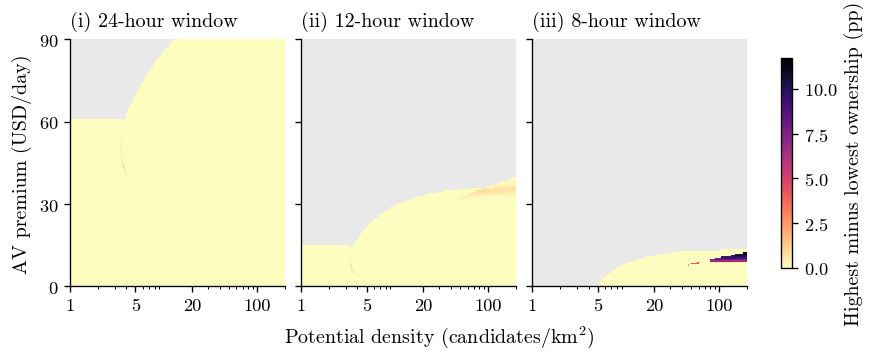}
    \caption{Difference between the highest and lowest ownership shares among detected active equilibria, in percentage points, under the 24-, 12-, and 8-hour windows.}
    \label{fig:ownership_gap}
\end{figure}

\Cref{fig:ownership_gap} reports the difference between the largest and smallest ownership shares among detected active equilibria at each market configuration.

\subsection{{Ownership sensitivity to market parameters}} \label{app:sens}
\begin{figure}[ht]
\centering
\includegraphics[width=0.6\linewidth]{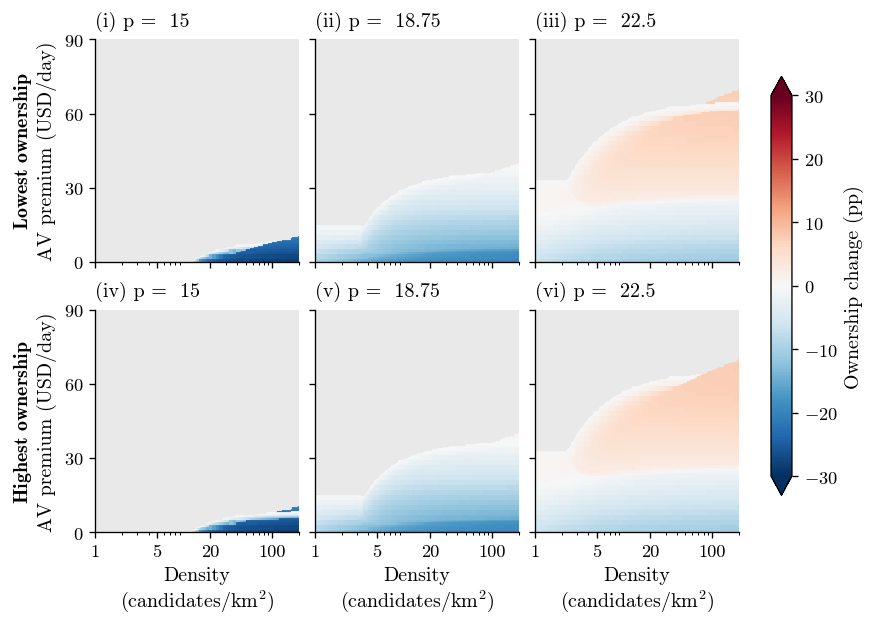}
\caption{Ownership change relative to no trade, in percentage points, for three fares $p$. Top row: lowest-ownership active equilibrium. Bottom row: highest-ownership active equilibrium.}
\label{fig:app_fare}
\end{figure}

\begin{figure}[ht]
\centering
\includegraphics[width=0.6\linewidth]{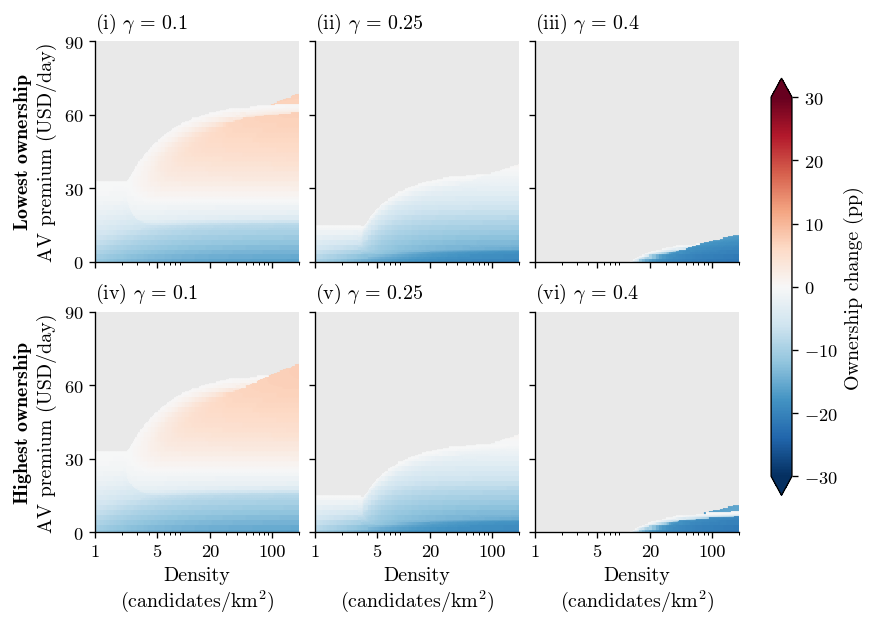}
\caption{Ownership change relative to no trade for three commissions $\gamma$, with rows as in \Cref{fig:app_fare}.}
\label{fig:app_commission}
\end{figure}

\begin{figure}[ht]
\centering
\includegraphics[width=0.6\linewidth]{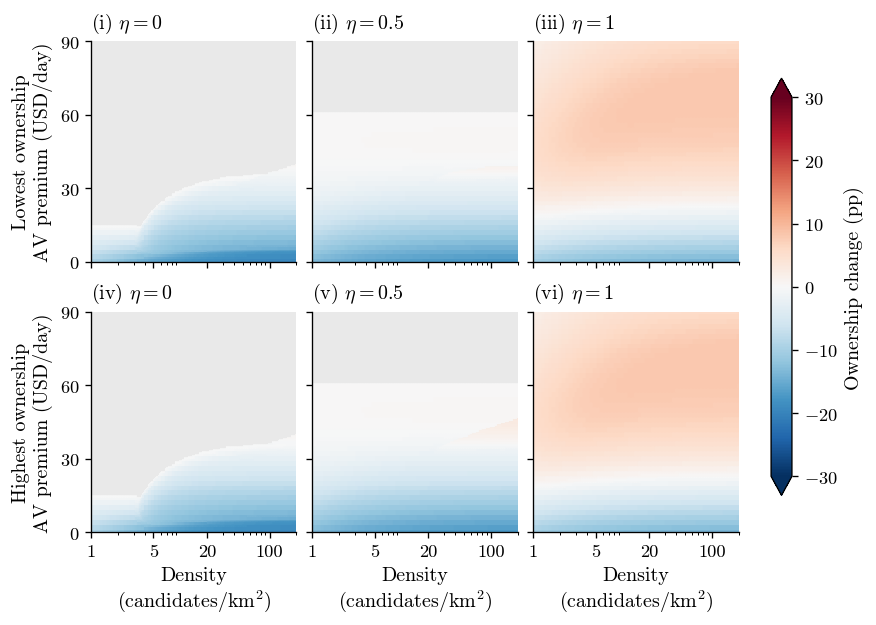}
\caption{{Ownership change relative to no trade for three levels of recall protection $\eta$, with rows as in \Cref{fig:app_fare}. Full protection ($\eta=1$) removes the access penalty, and sharing then increases ownership in high-premium markets.}}
\label{fig:app_recall}
\end{figure}

This appendix reports the full sweeps behind Case Study~II. All figures use the 12-hour window and report, for each parameter value, the ownership change relative to no trade of the lowest-ownership (top rows) and of the highest-ownership (bottom rows) active equilibrium over $(\lambda,\Cav)$. Gray indicates that no active equilibrium is detected. In the fare and commission sweeps, the middle columns correspond to the baseline of \Cref{tab:parameters}. \Cref{fig:app_fare,fig:app_commission} show that a higher fare and a lower commission enlarge the set of active markets and create a region of dense, high-premium markets in which sharing increases ownership. At the lowest fare and the highest commission, active markets survive only at high density and low premium. \Cref{fig:app_recall} shows the role of recall protection. Since protection rescales the access-failure weight from $\mu$ to $(1-\eta)\mu$ (\Cref{subsec:cs2}), this sweep is equivalent to a sweep over $\mu$, and full protection corresponds to $\mu=0$.

\end{document}